\documentclass[10pt,journal,compsoc]{IEEEtran}

\ifCLASSOPTIONcompsoc
  \usepackage[nocompress]{cite}
\else
  \usepackage{cite}
\fi

\usepackage{graphicx}
\usepackage{comment}
\usepackage{amsmath}

\usepackage{bbding}

\usepackage{color}
\usepackage[normalem]{ulem}

\usepackage[colorlinks,linkcolor=green]{hyperref}
\ifCLASSINFOpdf
\else
\fi
\usepackage{url}

\begin{document}
%
\title{MFDL: A Multicarrier Fresnel Penetration Model based Device-Free Localization System leveraging Commodity Wi-Fi Cards}
%
%
%
%

\author{Hao~Wang,~\IEEEmembership{Student Member,~IEEE,}
        Daqing~Zhang, \IEEEmembership{Member,~IEEE,} ~Kai~Niu, \\~Qin~Lv, ~Yuanhuai~Liu, ~Dan~Wu, ~Ruiyang~Gao and Bing Xie 
\IEEEcompsocitemizethanks{\IEEEcompsocthanksitem H. Wang, D. Q. Zhang, K. Niu, D. Wu, R. Y. Gao and B. Xie are with the Key Laboratory of High Confidence Software Technologies, Ministry of Education, and School of Electronics Engineering and Computer Science, Peking University; Y. H. Liu, academy for advanced interdisciplinary studies, Peking University; Q. Lv, Department of Computer Science, University of Colorado Boulder. \protect
E-mail:\{wanghao,dqzhang\}@sei.pku.edu.cn; qin.lv@colorado.edu;
 \{xjtunk, gry,dan,yunhuai.liu,xiebing\}@pku.edu.cn \protect}
\thanks{Manuscript submitted in July, 2017.}}

\IEEEtitleabstractindextext{%
\begin{abstract}
Device-free localization plays an important role in many ubiquitous applications. Among the different technologies proposed, Wi-Fi based technology using commercial devices has attracted much attention due to its low cost, ease of deployment, and high potential for accurate localization. Existing solutions use either fingerprints that require labor-intensive radio-map survey and updates, or models constructed from empirical studies with dense deployment of Wi-Fi transceivers. In this work, we explore the Fresnel Zone Theory in physics and propose a generic Fresnel Penetration Model (FPM), which reveals the linear relationship between specific Fresnel zones and multicarrier Fresnel phase difference, along with the Fresnel phase offset caused by static multipath environments. We validate FPM in both outdoor and complex indoor environments. Furthermore, we design a multicarrier FPM based device-free localization system (MFDL),  which overcomes a number of practical challenges, particularly the Fresnel phase difference estimation and phase offset calibration in multipath-rich indoor environments. Extensive experimental results show that compared with the state-of-the-art work (LiFS), our MFDL system achieves better localization accuracy with much fewer number of Wi-Fi transceivers. Specifically, using only three transceivers, the median localization error of MFDL is as low as 45$cm$ in an outdoor environment of 36$m^2$, and 55$cm$ in indoor settings of 25$m^2$. Increasing the number of transceivers to four allows us to achieve 75$cm$ median localization error in a 72$m^2$ indoor area, compared with the 1.1$m$ median localization error achieved by LiFS using 11 transceivers in a 70$m^2$ area.
\end{abstract}

}

\maketitle

\IEEEdisplaynontitleabstractindextext

%
\IEEEpeerreviewmaketitle

\IEEEraisesectionheading{\section{Introduction}\label{sec:introduction}}

%
%
%
%
\IEEEPARstart{I}{ndoor} localization is a fundamental building block for many ubiquitous computing applications in the real-world such as indoor navigation, assistive living, and context-aware computing in general. 
In recent years, device-free localization has attracted a lot of attention from both academia and industry \cite{wilson2011see,wilson2012fade,lin2007doppler,mao2011ilight,wagner2012passive,ruan2014tagtrack,zhang2007rf,xiao2013pilot,abdel2013monophy,wang2016lifs}. Since no additional device needs to be carried by the target, device-free localization can locate a target in a non-intrusive and privacy-preserving manner, which is particularly desirable in many real-world applications such as intrusion detection \cite{wilson2011see,wu2012fila}, elderly care \cite{miskelly2001assistive}, and  patient tracking \cite{gao2006vital}.

Many technologies have been proposed for indoor localizations, such as video \cite{ma2012reliable}, laser \cite{zhang2000line}, infrared \cite{kemper2010passive}, and pressure \cite{nukaya2012noninvasive}. Among all these technologies, Wi-Fi is one of the most promising approaches because of its ubiquity and easy deployment. Using Commercial Off-The-Shelf (COTS) wireless routers, Wi-Fi has been widely used for the last-mile connection of mobile devices. A cost-effective localization service can be easily offered by simply augmenting these routers. In addition, unlike laser, video, or infrared based solutions, the Wi-Fi technology has no special requirement on lighting, temperature, or other infrastructural support, making it particularly suitable for diverse environments.

Wi-Fi based device-free localization is based on a simple observation. In a static environment (either open space or complex indoor scenarios), the radio signals between two transceivers (COTS routers in our case) are fairly stable. When moving objects appear nearby, the signals can vary substantially. By measuring such variations, the mobile objects can be perceived and their locations can be inferred. These radio signal variations can be measured by the coarse-grained Radio Signal Strength Index (RSSI), as has been done in traditional approaches \cite{zhang2007rf,xu2013scpl,seifeldin2013nuzzer,youssef2007challenges}. Many recent works have adopted the more comprehensive Channel State Information (CSI), which includes finer-grained signal amplitude and phase information on each subcarrier \cite{sen2011precise,sen2012you,xiao2012fifs,abdel2013monophy,xiao2013pilot,kotaru2015spotfi,vasisht2016decimeter,li2016dynamic}.

Existing Wi-Fi based device-free localization methods can be roughly classified as fingerprinting-based and model-based. Fingerprint-based approaches \cite{xiao2013pilot} collect CSI measurement (or RSSI in earlier works \cite{youssef2007challenges,seifeldin2013nuzzer}) at each location and build a CSI site-map. When targets are present, this site-map is consulted to obtain the location of the target. Fingerprint-based approaches are widely challenged by its substantial labor work for site-map survey and update. Model-based approaches attempt to build a unified model to quantify the relationship between CSI measurements and the target locations. For example, Zhang et.al. \cite{zhang2007rf} proposed a SVR model that associates target locations with a triangle setting of the transceivers. FILA \cite{wu2012fila} uses path loss model to estimate the distance to known APs, location is then calculated based on trilateration. A more recent work LiFS \cite{wang2016lifs} applies the diffraction fading model and power fading model \cite{molisch2012wireless} to select a number of ``clean'' subcarriers with less multipath effects. All these works, however, build their models based on empirical studies. They employ wireless links as the basic sensing units. As such, these works require dense deployment of transceivers in order to achieve fine-grained localization. For instance, LiFS achieves 1.1m localization accuracy with 11 transceivers in a 70$m^2$ indoor area. Apparently, there is a lack of a fine-grained model directly linking the CSI measurements to a moving person's location, developing such a fine-grained model is of great value for accurate device-free localization.

\begin{figure*}[!t]
	\centering
	\includegraphics[width=6.5in]{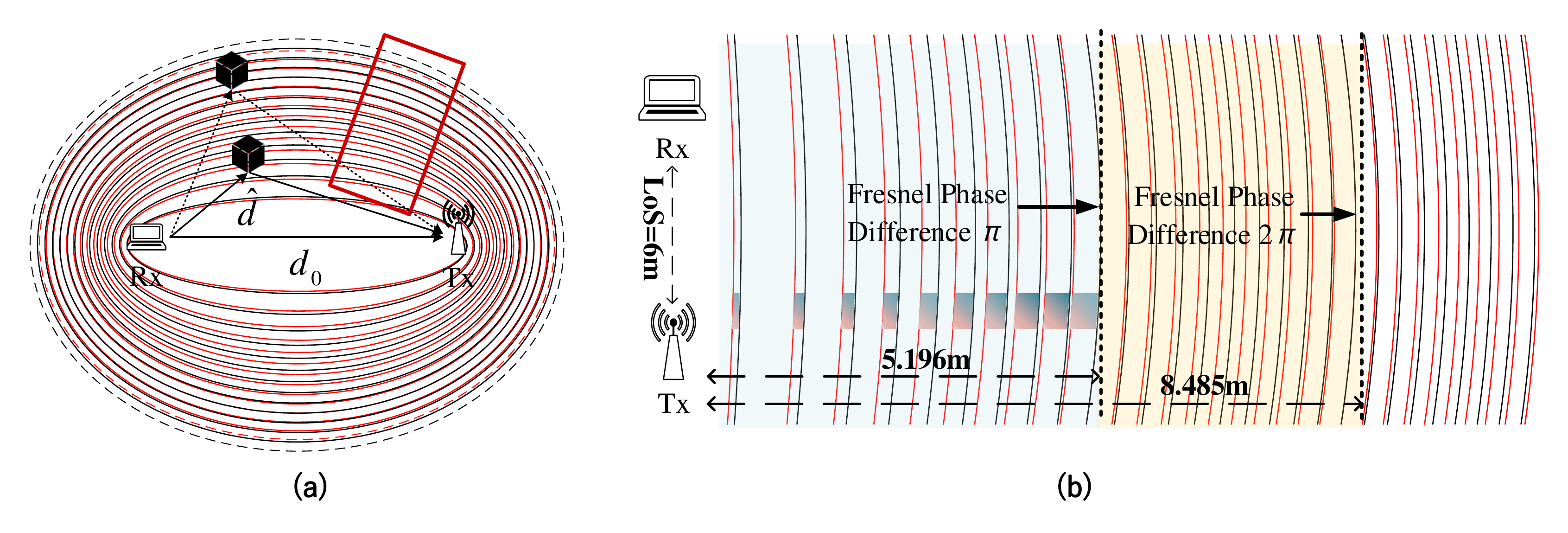}
	\caption{Conceptual illustration of (a) the Fresnel zones and (b) the zoom-in view of multicarrier Fresnel zones for a specific setting}
	\label{Fig11}
\end{figure*}

\begin{figure}[!t]
	\centering
	\includegraphics[width=1.9in]{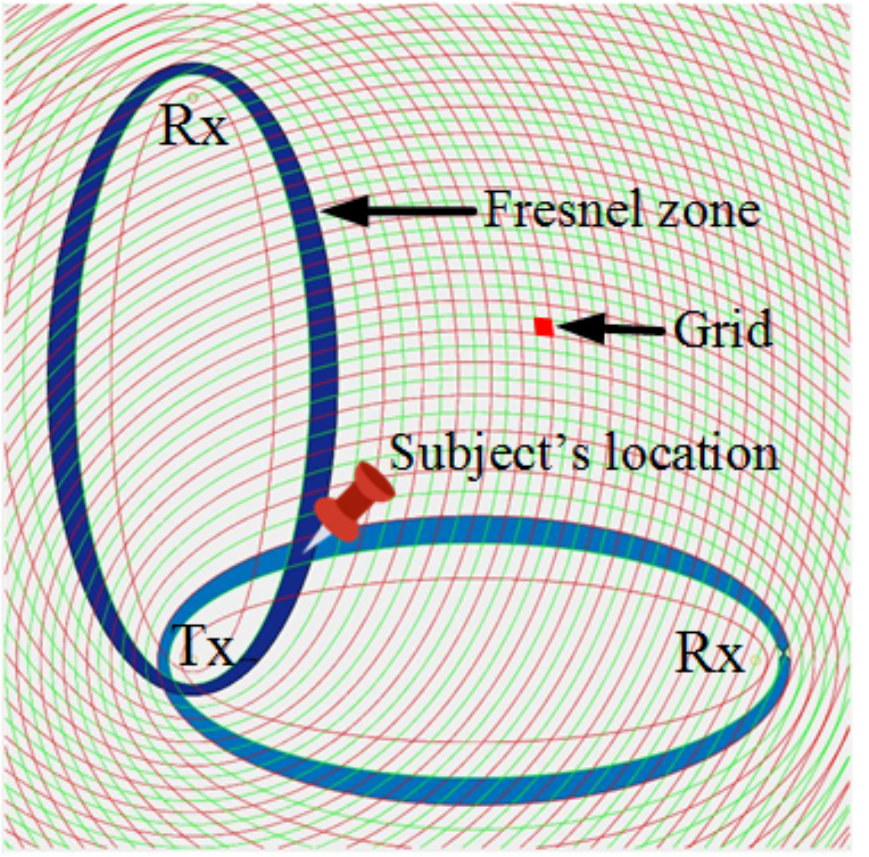}
	\caption{Basic idea of 2-dimenstion multicarrier FPM based device-free localization system (MFDL).}
	\label{Fig21}
\end{figure}

To this end, in this work, we would like to leverage the properties of Multicarrier Fresnel zones uncovered in \cite{zhang2017toward} to develop a decimeter-scale localization model. As Wi-Fi 802.11n+ specification leverages an orthogonal frequency-division multiplexing (OFDM) based transmission scheme, it divides the whole bandwidth into multiple subcarriers with different frequencies. Multiple Fresnel zones are thus formed around the transmitter (Tx) and receiver (Rx) antennas according to their wavelengths. These multicarrier Fresnel zones share the same foci and take the shape of ellipsoids with different sizes: a subcarrier with a shorter wavelength has smaller ellipsoid, as shown in Fig. \ref{Fig11}(a). In the inner Fresnel zones, the same layer of Fresnel ellipsoids of different subcarriers almost overlap with one another. With the number of Fresnel zones increasing as shown in Fig. 1(b), the gap between a pair of Fresnel zones , which we call the Fresnel Phase difference, keeps increasing monotonically until the boundary of the (i+1)th layer of Fresnel zone of smaller wavelength catches up with that of the ith Fresnel zone with larger wavelength. Thus by placing a pair of WiFi transceivers apart with sufficient distance, the monotonic relationship between the Fresnel phase difference and one's position in Fresnel zones is upheld. If it's possible to map the Fresnel Phase difference to its position in the Fresnel zones of a pair of transceivers (i.e., map a subject's location to a cluster of adjacent Fresnel zones (rings) as shown in Fig. \ref{Fig21}, then we can locate a moving subject by finding the intersection area between two Fresnel zone rings produced by two pairs of WiFi transceivers as shown in Fig. \ref{Fig21}. Considering the size of one Fresnel zone being around a few centimeters for 5GHz WiFi, decimeter-level even centimeter-level localization accuracy could be expected if the model mapping can be done accurately in real-world environments.

However, several challenges exist to achieve robust and precise localization. First, how to obtain the Fresnel Phase difference accurately and correlate it with the right Fresnel zones mathematically in noisy environments. Second, how to understand and characterize the distorted Multicarrier Fresnel zones in the indoor multipath-rich environment, and how to restore the correlation between the Fresnel Phase difference and a moving subject's position in Fresnel zones in such complex environment. Third, on one hand, human body is not a perfect radio reflector and radio signals are absorbed resulting in distortion in the raw CSI measurements. On the other hand, any pair of WiFi subcarriers can be used in theory to compute the Fresnel phase difference for localization, how to choose among the numerous subcarrier-pairs to obtain precise Fresnel phase difference and Fresnel zone position, with robust and accurate localization performance.


In order to address the above challenges, we start by introducing the properties of the basic Fresnel zone model for a single subcarrier and pointing out that it's not sufficient for device-free localization.
By analyzing the properties of Multicarrier Fresnel zones, and the induced distortion caused by indoor multipath in real-world environment, we propose a generic and fine-grained model called the {\bf Fresnel Penetration Model (FPM)} which correlates the Fresnel Phase difference with the location of Fresnel zones mathematically in both free and indoor space. Based on the proposed FPM model, we develop a novel and accurate device-free localization system called MFDL, which contains three key components, namely Fresnel phase difference estimation, Phase offset calibration and Model Fitting, to tackle the practical issues in real-world environments.

The main contributions of our work are as follows: 

1) To the best of our knowledge, FPM is the first fine-grained localization model that can directly quantify the relationship between the target location and WiFi CSI measurements in both open space and multipath-rich indoor environments. In particular, we first reveal the linear relationship between the specific Fresnel zone numbers that a moving object resides in and the Fresnel zone phase difference as well as how static multipath environments affect the Fresnel phase offset.

2) We conduct intensive empirical studies to validate FPM in open space and real-world indoor environments. Experimental results show high consistence with the modeling results. The median localization error is only 6cm in the open space and 13cm in the indoor environments with a metal plate as reflector.

3) Based on FPM, we design and implement MFDL, a novel multicarrier FPM based device-free localization system in which we overcome a series of challenges, particularly the Fresnel phase difference estimation and phase offset calibration in multipath-rich indoor environments.

4) We conduct comprehensive field studies to evaluate the performance of MFDL. Experimental results show that using 3 WiFi transceivers, MFDL achieves a median localization error of 45cm in an open area of 36$m^2$, and 55cm in indoor areas of 25$m^2$. Increasing the number of transceivers to four allows us to achieve 75cm median localization error in a 72$m^2$ indoor area,  compared with the 1.1m median error achieved by the state-of-the-art model-based localization system LiFS using 11 transceivers in a 70$m^2$ area.

\section{Related Work}
\label{sec:related} 

		Recent years have seen growing research interests in COTS Wi-Fi based device-free localization. Firstly, \textbf{compared with non-RF based technologies} such as video \cite{ma2012reliable}, infrared \cite{kemper2010passive}, visible light \cite{li2015human} or pressure \cite{nukaya2012noninvasive}, Wi-Fi based technology demonstrates great potentials in terms of ubiquity and ease of deployment. It requires no additional or customized hardware, is insensitive to light or temperature, and provides better privacy protection to locate human targets without leaking potentially sensitive information such as face or weight or audio recordings. Secondly, \textbf{compared with other RF-based prior works} such as Wi-Vi \cite{adib2013see}, WiDeo \cite{joshi2015wideo}, WiTrack \cite{adib2015multi}, mTrack \cite{wei2015mtrack}, and Tadar \cite{yang2015see}, Wi-Fi based technology does not require special hardware such as USRP to send out special radio waves or customized RFID reader arrays \cite{yang2015see}. Leveraging COTS Wi-Fi routers, it can be easily applied by augmenting existing Wi-Fi infrastructures.
		
		Existing COTS Wi-Fi based device-free localization works can be further classified as fingerprint-based or model-based. \textbf{Fingerprint-based approaches} built a radio-map and refer to this map when localizing the target. Early works utilized RSS information such as Nuzzer \cite{seifeldin2013nuzzer} and Ichnaea \cite{saeed2014ichnaea}. These works are limited by the inherently corse-grained property of RSS values. For example, Ichnaea achieved a median accuracy of 2.5m with high density deployment. Recent research works have utilized the finer-grained CSI information, which simultaneously captures amplitude and phase across all individual subcarriers. For example, E-eye \cite{wang2014eyes} built a CSI amplitude map to determine the target moving trajectory. Fingerprint-based approaches need extensive labor work for radio-map survey and update when the environment changes. Although crowdsourcing \cite{wu2013will,rai2012zee}, 3D ray tracing\cite{eleryan2011synthetic,aly2013new}, and model transformation methods \cite{ohara2015transferring} have been proposed to greatly reduce the overhead, the cost is still prohibitively high and the localization accuracy remains at meter level. \textbf{Model-based approaches} establish mathematical relationships between RF measurements and target locations. SVR \cite{zhang2007rf} employed a Support Vector Model, which needed high-density deployment of transceivers. MaTrack \cite{li2016dynamic} used CSI phase information to build an Angle-of-Arrival (AoA) model for the target. And the state-of-art LiFS applied the diffraction fading model and power fading model to empirically select subcarriers that fit the model. All these works, however, worked well for target objects that are close to the LoS (Line-of-Sight) path of transceivers, and thus high accuracy depends on high-density deployment. For example, LiFS achieves 1.1m median error with 4 APs and 7 clients in a 70$m^{2}$ indoor room. In comparison, our approach is based on precise physical modeling. It not only reduces the localization error to decimeter level,   but also requires much fewer number of transceivers. 
		
		Our work is based on Fresnel zones, and here we review prior works w.r.t Wi-Fi Fresnel zone based human sensing. The notion of ``Fresnel zone'' was first proposed in the early nineteenth century \cite{jenkins1957fundamentals} and was explicitly introduced by \cite{wang2016human} and \cite{wu2016widir} as the theoretical basis for indoor human sensing with Channel State Information (CSI) of Wi-Fi signals, and the key findings with further implications are summarized in \cite{zhang2017toward}. Even though prior work \cite{zhang2017toward} reveals the relationship among the Fresnel phase difference, the signal propagation path difference and the subcarrier frequency difference, it neither investigate the impact of the multipaths on the Fresnel phase difference, nor proposing any solutions to Fresnel phase difference estimation for localization exploiting different properties and insights. In contrast, in this work, we first explicitly state the linear relationship between the subject's residing Fresnel zone and the Fresnel phase difference  for a fixing pair of subcarriers; second, we propose to compute the subject's location by finding the intersection area between two Fresnel zone rings produced by two pairs of Wi-Fi and robustly estimate the Fresnel phase difference for localization purpose exploiting various properties and insights of FPM; More importantly, we reveal and characterize how the static multipaths affect the Fresnel phase difference leading to a Fresnel phase offset in indoor environments and how this the phase offset can be compensated with a simple calibration mechanism to increase the localization accuracy. As a result, our system achieves the best-ever  decimeter-level localization accuracy. \\

\section{Background}
In this section, we briefly introduce the Wi-Fi CSI measurements which reveal the fine-grained channel information at the scale of OFDM multi-subcarriers, then present the basics of the Fresnel Zone Theory as the foundation of this work.

\subsection{WiFi CSI}
Wi-Fi 802.11n+ specification leverages the OFDM based transmission scheme, which divides the whole bandwidth into multiple subcarriers with different frequencies. Commodity WiFi devices report CSI at the granularity of OFDM subcarrier level. For example, the Intel 5300 wireless NIC reports 30 CSI values and each value corresponds to one CSI subcarrier. These CSI subcarriers are spaced in 2$\times$0.3125MHz=0.625MHz and 4$\times$0.3125MHz=1.25MHz, for 20MHz and 40MHz bandwidth, respectively.

\subsection{Fresnel Zone}
The concept of Fresnel zone originated from the research on the interference and diffraction of light in the early nineteenth century \cite{jenkins1957fundamentals}. In the context of radio propagation in a 2-dimensional plane, Fresnel zones refer to the concentric ellipses with foci in a pair of transceivers. Assume $T_x$ and $R_x$ are two transceivers with certain height (as shown in Fig. \ref{Fig11}(a)), for a given radio signal with wavelength $\lambda$, the Fresnel zones containing $n$ ellipses can be constructed by ensuring:
\begin{displaymath}
|T_xQ_n|+|Q_nR_x|-|T_xR_x|=n\lambda/2\
\end{displaymath}
where $Q_n$ is a point on the $n$th ellipse while the ellipses themselves are the Fresnel zone boundaries. The innermost ellipse is defined as the $1$st Fresnel zone, the elliptical annuli between the first ellipse and the second is defined as the $2$nd Fresnel zone, and the $n$th Fresnel zone corresponds to the elliptical annuli between the $(n-1)$th and $n$th ellipses. As revealed in \cite{zhang2017toward,wang2016human}, an object in the field will produce a reflected signal and the received signal in Rx is a linear combination of the reflected signal and signal via LoS. When the object passes through a series of Fresnel zones (as shown in Fig. \ref{Fig11}(a)), the receiving signal shows a continuous sinusoidal-like wave, with peaks and valleys generated by crossing the boundaries.

\section{Fresnel Penetration Model}
In this Section, we explore the Fresnel Zone Theory to establish the connection between the CSI measurements and the object resided Fresnel Zones. 
We start from the simplest case of single subcarrier and single NLoS (Non-Line-of-Sight) path reflected by the object (Section 4.1) to introduce the notion of Fresnel phase and reveal its limitations for localization. 
Then we extend it to multi-subcarrier Fresnel zones (Section 4.2) and quantify the linear relationship between Fresnel phase difference of subcarrier-pairs and the resided Fresnel zones in theory. 
After that, we study the most challenging but realistic scenario with multiple NLoS paths (Section 4.3). By characterizing how the multi-path distorts multicarrier Fresnel zones, we demonstrate how to restore the correlation between the Fresnel Phase difference and the resided Fresnel zones in multipath rich settings.
In the last, we verify our FPM model with a steel plate as reflector in open space and real-world indoor environments (Section 4.4).

\subsection{FPM with Single Subcarrier}
As shown in Fig.\ref{Fig11} (a), assume a pair of transceivers and an object are the only objects in the field. When the object moves, it will pass through a series of Fresnel zones, then the receiving signal power $|H(\lambda,\hat d )|^2$ shows a continuous sinusoidal-like wave with peaks and valleys generated by crossing the Fresnel zone boundaries\cite{zhang2017toward,wang2016human}:

\begin{equation}\label{e4}
\begin{split}
|H(\lambda,\hat d )|^2 = |H_s(\lambda)|^2 +|H_d(\hat d,\lambda)|^2 +\\ 2|H_s(\lambda)||H_d(\hat d,\lambda)|cos\varphi(\lambda, \hat d)
\end{split}
\end{equation}
where $\hat d$ is the path length of the reflected signal from the moving object and $\lambda$ is the wavelength of the subcarrier. The static vector $H_s(\lambda)$ is the LoS signal while the dynamic vector $H_d(\hat d,\lambda)$ is introduced by the reflected signal from the moving object as shown in Fig.\ref{Fig11} (a). The term $\varphi(\lambda, \hat d)$, which we name as the \textbf{Fresnel Phase}, is the phase difference between the static vector $H_s(\lambda)$ and dynamic vector $H_d(\hat d,\lambda)$ that can be further represented as:
\begin{equation}\label{e5}
\varphi(\lambda, \hat d) = 2\pi (\hat d - {d_0})/\lambda
\end{equation}
where $d_0$ is the length of LoS signal. As $d_0$ and $\lambda$ are constant, the Fresnel phase varies as the reflected path length changes as a result of object movement. Apparently, according to Equation \ref{e4}, it can be seen that the CSI power $|H(\lambda,\hat d )|^2$ for a single subcarrier is a time-varying signal, having no obvious correlation with the moving object's location. In another word, Equation \ref{e4} can only tell us that the object is crossing the Fresnel zones, but cannot inform which Fresnel zones the object resides in.

\subsection{FPM with Multiple Subcarriers}
Based on the discussion above, we understand the limitation of using Fresnel zone of a single subcarrier wave for localization. As we have mentioned in Section 3.1, commodity WiFi devices report CSI at the granularity of OFDM subcarrier level, Multiple Fresnel zones are thus formed independently around the pair of transceivers. In a 2-dimensional plane, these multicarrier Fresnel zones share the same foci and take the shape of ellipses with different sizes: a subcarrier with shorter wavelength has smaller ellipses, as shown in Fig. \ref{Fig11}(a). Specifically, in the inner Fresnel zones, the corresponding Fresnel zone boundaries of different subcarriers almost overlap with one another. However, as the number of Fresnel zone increases, the gap between a pair of Fresnel zones of two fixed subcarriers keeps increasing as shown in Fig. \ref{Fig11}(b), until the boundary of the (i+1)th Fresnel zone of the subcarrier with the smaller wavelength catches up with that of the ith Fresnel zone with larger wavelength. 
For instance, when the distance between the  transceivers is 6m, for two subcarriers with frequency 5.745GHz and 5.770GHz, the gap is monotonic before the 230th Fresnel zone boundary of subcarrier 5.770GHz catches up with 231th Fresnel zone boundary of the subcarrier 5.745GHz, at the location which is approximately 5.196m from the LoS between two transceivers. This inspires us that if we can quantify this gap from the CSI measurements and model the correlation between the gap and the object's location in Fresnel zones, we can infer the Fresnel zones the object resides in. Actually, quantifying this gap is equivalent to calculating the difference between the two Fresnel phases (see Equation \ref{e5}) of two subcarriers:
\begin{equation}\label{e6}
\begin{array}{lll}
\Delta {\varphi _{ab}}(\hat d) &=& \varphi ({\lambda _a},\hat d) - \varphi ({\lambda _b},\hat d)\\
&=& 2\pi (\hat d - {d_0}) \Delta f /c\\
&=& \lambda _a m_a \pi \Delta f/c = \lambda _b m_b \pi \Delta f/c 
\end{array}
\end{equation}
where $\Delta f$ is the frequency gap between the two subcarriers $\lambda_a$ and $\lambda_b$; $c$ is the light speed; $m_a$ and $m_b$ is the sequence number of the resided Fresnel zones w.r.t the two subcarriers  $\lambda_a$ and $\lambda_b$.

\bf{Fresnel Penetration Model (FPM) :} \rm Given two subcarriers $\lambda_a$ and $\lambda_b$, the \textbf{Fresnel Phase Difference}, denoted as $\Delta \varphi_{ab}(\hat d)$, has a linear relationship with the resided Fresnel zones in theory. By measuring the Fresnel phase difference from the raw CSI values, the moving object resided Fresnel Zones can be inferred; by finding the intersection area between two Fresnel zone rings produced by two pairs of WiFi transceivers as shown in Fig. \ref{Fig21}, the moving object's position can be located.

\begin{figure}[!t]
	\centering
	\includegraphics[width=3.5in]{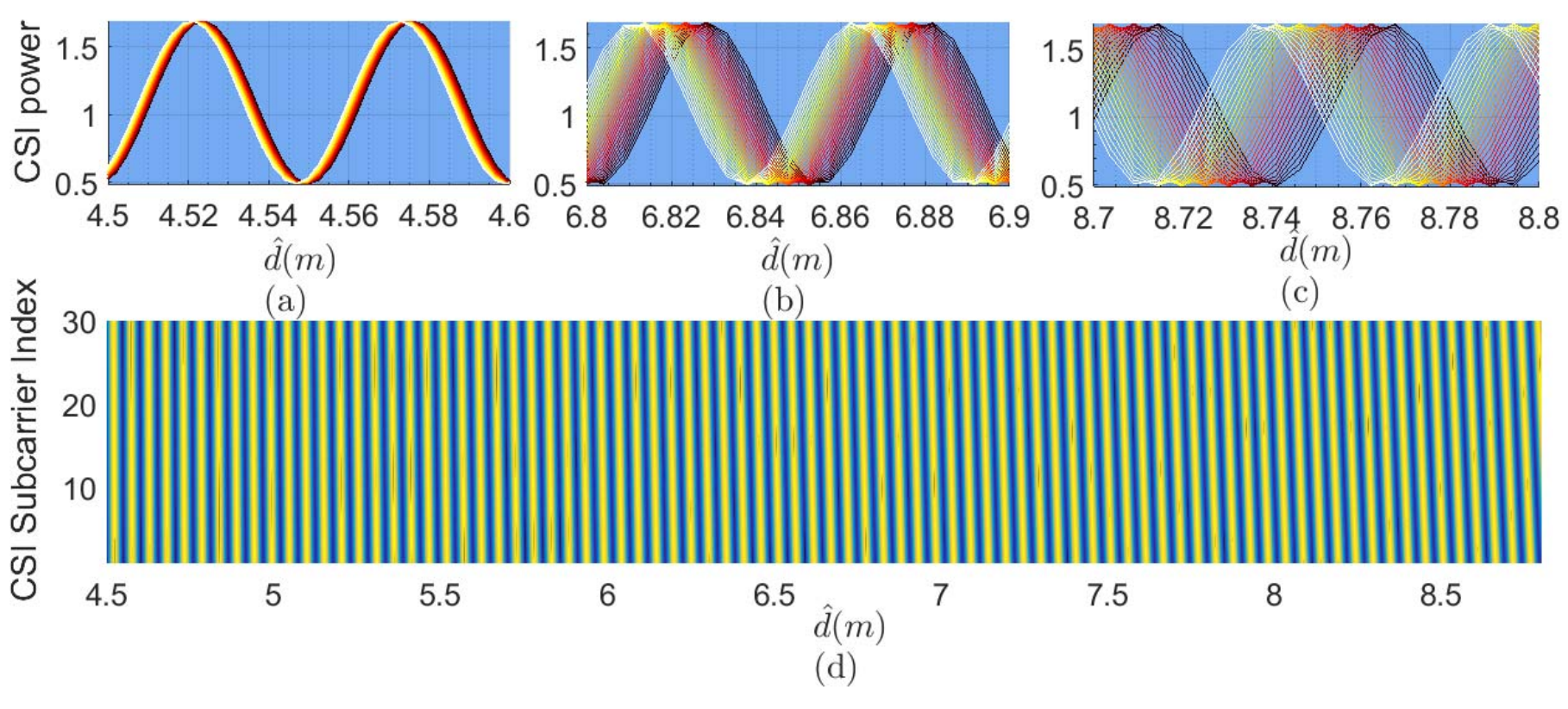}
	\caption{CSI amplitudes and heatmap in a simulated free space with one moving object}
	\label{Fig422}
\end{figure}

Fig. \ref{Fig422} shows CSI amplitudes of multi-subcarriers  in a simulated free space with one object. The number of subcarriers is set to 30, as supported by intel 5300 NIC \cite{halperin2011tool}. The central frequency is set to 5.745GHz with 40MHz bandwidth. The two transceivers are set $d_0=4m$ apart. The reflected path length $\hat d$ ranges from 4.5m to 8.8m, simulating a scenario where an object moves outward from the transceivers. Fig. \ref{Fig422}(a)-(c) depict the CSI amplitudes $|H(\hat d)|$ at a location close to the transceivers, in a middle range, and far from the transceivers, respectively. We can see that when $\hat d$ is small (Fig. \ref{Fig422}(a)),$\Delta \varphi _{ab}$  is small as well and the amplitude of different subcarriers are nearly overlapping with each other. While $\hat d$ increases, the Fresnel Phase Difference increases too. Fig. \ref{Fig422}(d) depicts the amplitude in a heatmap form where $x$-axis is $\hat d$, $y$-axis is the index of the subcarriers, and the color represents the amplitude of signal with dark color corresponding to higher values. On the one hand, if we zoom in the $x$-axis on a very small region (e.g., from 6.5-6.6), we can see almost straight lines. This is expected because according to Equation \ref{e6}, if we fix $\hat d$ as constant, the Fresnel phase difference $\Delta \varphi _{ab}$ has a linear relation w.r.t the frequency gap $\Delta f$; On the other hand, if we zoom in the $y$-axis on a subcarrier pair (e.g., subcarrier index 1 and 20), we can see that the slopes of the straight lines become notable when $\hat d$ is increasing. This is also expected because according to Equation \ref{e6}, if we fix $\Delta f$ as constant, the Fresnel phase difference $\Delta \varphi _{ab}$ has a linear relation w.r.t the reflected path length $\hat d$.
\begin{figure}[!t]
	\centering
	\includegraphics[width=3.5in]{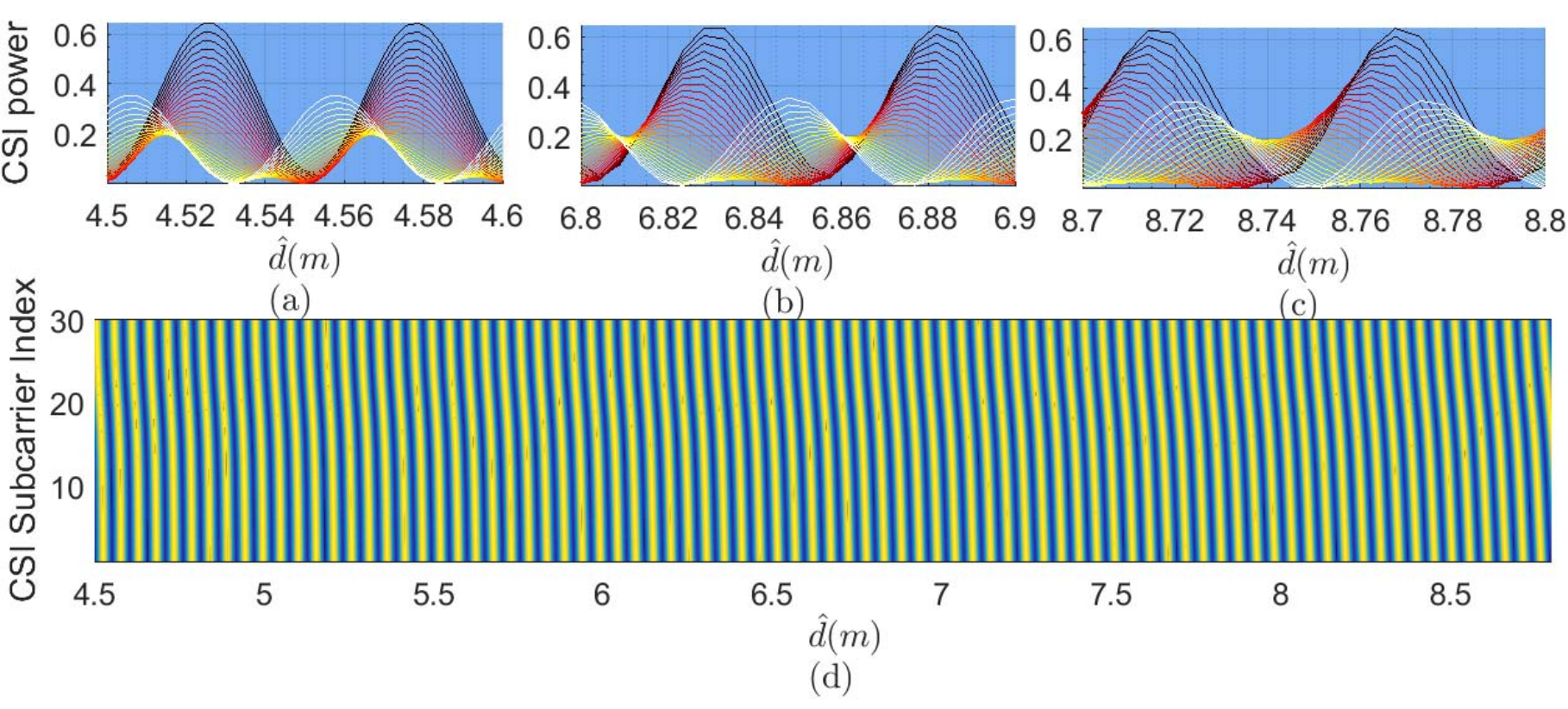}
	\caption{CSI amplitudes and heatmap in a simulated multipath rich environment with one moving object}
	\label{Fig43}
\end{figure}

\begin{figure*}[!t]
	\centering
	\includegraphics[width=6.7in]{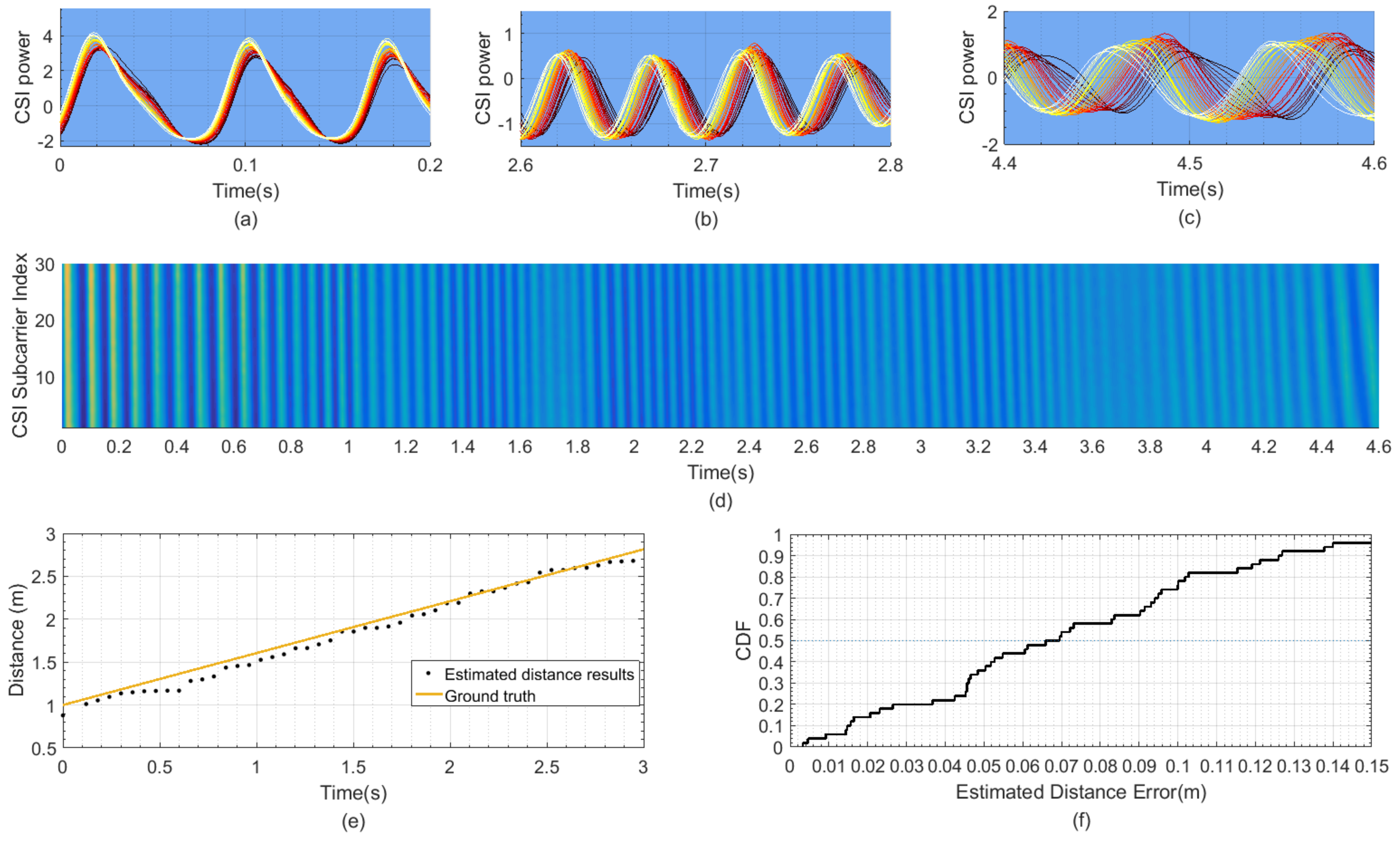}
	\caption{Measured CSI amplitudes, heatmap and localization error for a perfect reflector in an outdoor open space}
	\label{Fig441}
\end{figure*}
\subsection{FPM in Multipath Rich Environment}
In a real indoor environment, there could be multiple radio propagation paths from the transmitter to the receiver. The objective of this subsection is to review the basic FPM model proposed in previous section 3.2 and see if it still holds in multipath rich environment. If not, we need to understand the multipath induced distortion in the multicarrier Fresnel zones and restore the correlation between the Fresnel phase difference and a moving subject's position in Fresnel zones in indoor environment.

We divide all the paths into static and dynamic ones, then the Equation \ref{e4} which characterizes the  receiving signal power $|H(\lambda,\hat d )|^2$ should be re-written as follows \cite{wang2015understanding,tse2005fundamentals,wang2016human}:
\begin{equation}
\begin{split}
|H(\lambda,\hat d )|^2 = |\hat H_s(\lambda)|^2 +|H_d(\hat d,\lambda)|^2 + \\
2|\hat H_s(\lambda)||H_d(\hat d,\lambda)|cos \hat \varphi(\lambda, \hat d)
\end{split}
\end{equation}
where the static vector $\hat H_s(\lambda)$ is the sum of signals from all static paths while the dynamic vector $H_d(\hat d,\lambda)$ is introduced by the reflected signal from the moving object. The Fresnel phase $ \hat \varphi(\lambda, \hat d)$ indicates the phase difference between the static vector $H_s(\lambda)$ and dynamic vector $H_d(\hat d,\lambda)$ that can be further represented as:
\begin{equation}\label{e7}
\hat \varphi (\lambda, \hat d )=2\pi (\hat d-d_0)/\lambda 
+\varepsilon(\lambda)
\end{equation}
Compared with the Fresnel phase $ \varphi (\lambda, \hat d )$ in Equation \ref{e5}, we can see that in indoor environment, the Fresnel phase is distorted by the multipath with an unknown offset $\varepsilon(\lambda)$. Thus the Fresnel phase difference is also distorted with an unknown offset. Then the raw Fresnel phase difference in indoor multipath rich environment is as follows:
\begin{equation}\label{e8}
\begin{array}{lll}
\hat \Delta \varphi _{ab}&=& \hat\varphi(\lambda_a, \hat d)-\hat\varphi(\lambda_b,\hat d)\\
&=& \lambda _a m_a \pi \Delta f/c + {\varepsilon _{ab}}= \lambda _b m_b \pi \Delta f/c + {\varepsilon _{ab}}\\
&=& \Delta {{\varphi }_{ab}} + {\varepsilon _{ab}}
\end{array}
\end{equation}
where $\varepsilon_{ab} $ is a Fresnel Phase Offset associated with the two subcarriers introduced by static indoor multipath. As a consequence, we can see that in indoor multipath rich environment, the raw Fresnel Phase Difference still has a linear relation with the resided Fresnel zones except for an additional unknown offset which remains stable in static environments. In practice, we solve it by an offline calibration method presented in Section \ref{sec:eval}.

Fig. \ref{Fig43} shows the CSI amplitudes of multi-subcarriers  in a simulated multipath rich environment with the similar setting as in Fig. \ref{Fig422}. A fixed Phase Offset $\varepsilon$ is introduced for each pair of subcarriers as a result of being affected by the static multipaths. From the results we can find that on one hand, the key monotonic relation between $\Delta \varphi$ and $\hat d$ still holds. On the other hand, $\Delta \hat \varphi(\hat d)$ becomes a more complex function of $\hat d$ instead of linear relation as Equation \ref{e6}, leading to distorted curves in the heatmap shown in Fig. \ref{Fig43} (d). Notice that this distortion depends on the phase offset $\varepsilon$, and we will address this challenge in the localization system design (Section 4) using a Phase Offset Calibration method.
\subsection{Model Verification}
In this part, we conduct real experiments to verify Equation \ref{e6} and  \ref{e8} in FPM. Both outdoor and indoor environments are investigated and the results are reported as follows.
\subsubsection{Outdoor Evaluation with a Perfect Reflector}

The outdoor experiments are conducted in the open space as shown in Fig.~\ref{Fig621}(a). We employ a steel plate as the target object, as metals are perfect reflectors for radio signals. The two transceivers are placed $d_0=4m$ apart. We move the steel plate along the perpendicular bisector of the transceivers from 1m to 4m and measure the CSI amplitude. Fig.\ref{Fig441} shows the experimental results in an outdoor open space. We can see that the results obtained in real-world outdoor environment (as shown in Fig. \ref{Fig441} (a), (b), (c), (d)) are quite consistent  with that obtained by simulations (as shown in Fig. \ref{Fig422} (a), (b), (c), (d)). In particular,the curves in the heatmap Fig. \ref{Fig422}(d) are straight lines and the slopes are towards the higher frequency direction. This implies a simple linear relationship between $\hat d$ and the radio frequency. Fig. \ref{Fig441}(e) compares the estimated distances based on FPM with the ground truth, and the CDF of the estimated distance errors is depicted in Fig. \ref{Fig441}(f). In the outdoor open space with a metal plate as moving reflector, the median localization error using FPM is as low as 6cm, demonstrating the great potential of FPM for fine-grained moving object localization.
\begin{figure*}[!t]
	\centering
	\includegraphics[width=6.7in]{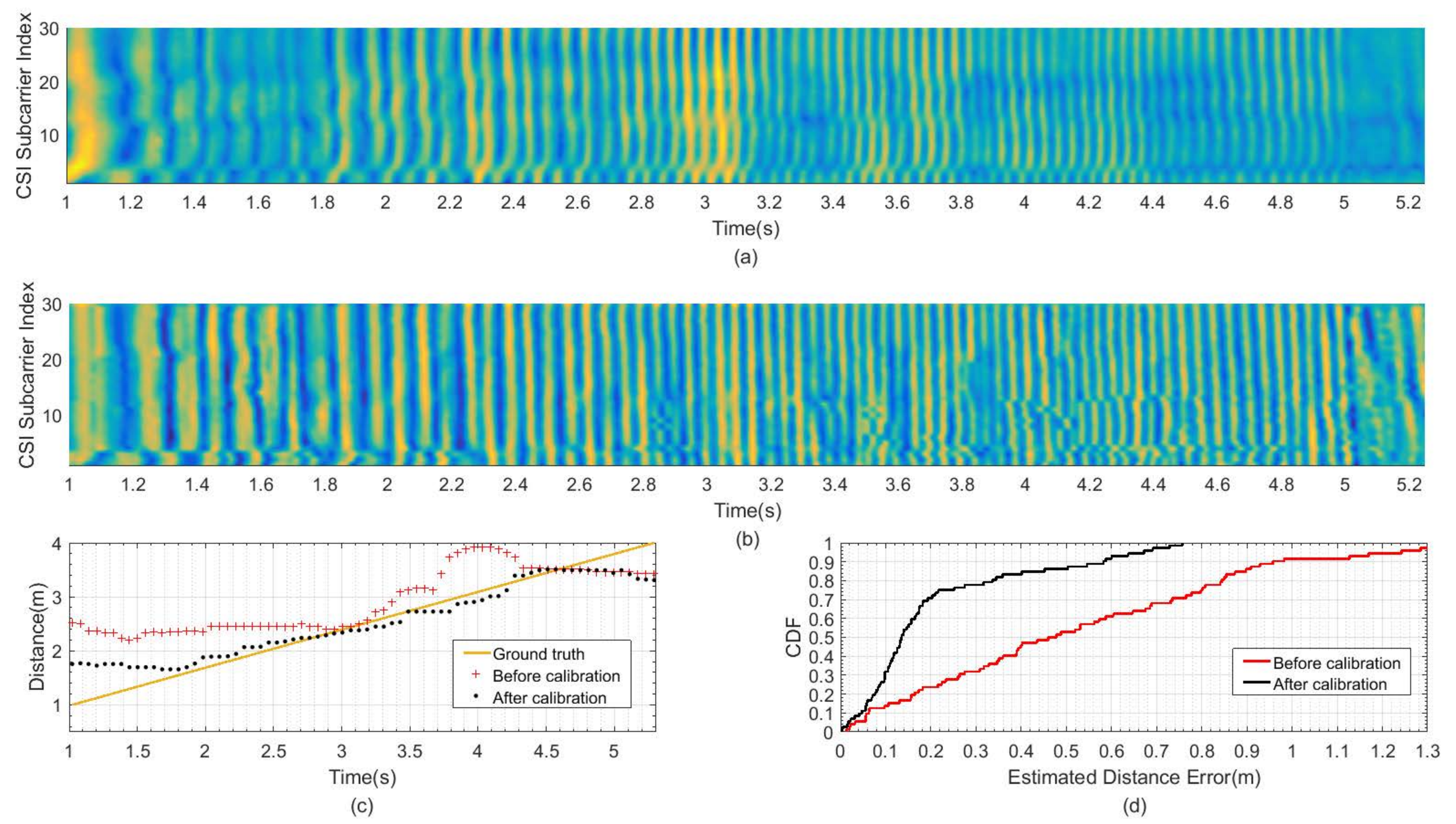}
	\caption{Heatmaps and localization errors with a perfect reflector in an indoor environment}
	\label{Fig442}
\end{figure*}
\subsubsection{Indoor Experiments with a Perfect Reflector}
We also perform the same field experiments in an indoor hall environment(see Fig. \ref{Fig621}(b)). Following the similar setting as outdoor environment, the two transceivers are also placed $d_0=4m$ apart and we move the steel plate along the perpendicular bisector of the transceivers from 1m to 4m and measure the CSI amplitude to verify the FPM model derived in Section 4.3. Not surprisingly, Fig \ref{Fig442} (a) shows that the curves in heatmap are distorted, different from those in outdoor environment in Fig. \ref{Fig441}(d). This is mainly due to the Fresnel Phase Offset $\varepsilon_{ab}$. Later in Section 4.4, we will introduce a Phase Offset Calibration algorithm. With such a calibration, the curves in Fig. \ref{Fig442}(a) are adjusted as straight lines, as shown in Fig. \ref{Fig442}(b). Fig. \ref{Fig442}(c) shows the estimated distance errors of FPM before and after the calibration and the CDF is plotted in Fig. \ref{Fig442}(d). It can be seen that the Phase Offset calibration effectively improves the localization accuracy and the median error reduces from 50cm to 13cm.

\section{MFDL System}
In this section, we present the detailed design of Multicarrier FPM based Device-Free Localization (MFDL) system. We first give an overview of the system architecture and identify the technical challenges in order to apply FPM in system design. We then describe each major functional component in detail.

\subsection{System Architecture}
In practical systems, we keep measuring the CSI amplitude $|H|$ at each subcarrier with the sampling rate 500 packets per second. The input of our system is thus a time series $|H(\lambda, t)|$ with the time interval $1/500=0.002$ seconds. 
As the object is moving,  the object reflected path length $\hat d(t)$ is a function of time $t$ too. The key issue in the system design is then to compute an accurate $\hat d(t)$ from $|H(\lambda,t)|$ for each time instant $t$ so that the real-time locations of the object can be inferred. The computation is conducted with respected to a sliding window $w$, involving the last $w$ values of $|H(t), t=t_{-w+1}, ..., t_0|$ from $t_0$. Later on, we will see the sliding window size $w$ is a key control parameter to the localization error.

\begin{figure}[!t]
	\centering
	\includegraphics[width=3.5in]{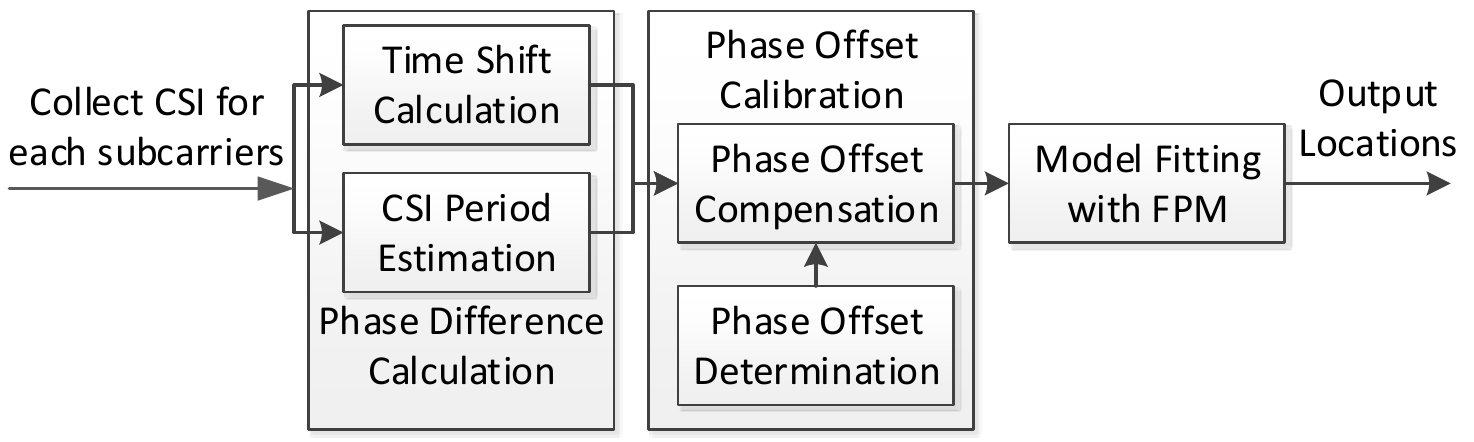}
	\caption{The MFDL localization system architecture}
	\label{Fig51}
\end{figure}

As shown in Fig. \ref{Fig51}, our system is composed of three main components, namely Fresnel Phase difference calculation, Phase Offset calibration, and model fitting with subcarrier-pairs.

The first step is to calculate the Fresnel Phase difference between any pair of subcarriers. It can be obtained by the time shift $\Delta T$  between amplitude time series $|H(\lambda, t)|$ of the two subcarrier, divided by the period $T$ of $|H(\lambda, t)|$ within the $w$ time window. When the time shift and CSI amplitude varying period obtained, the raw Fresnel Phase difference can be derived: $\Delta \hat \varphi_{ab}(\hat d)= \Delta T/T$ (Section 4.2).

Recall that in an indoor environment, there is a static Phase Offset $\varepsilon_{ab}$ contained in the raw Fresnel Phase difference. The second step is determine this offset and calibrate the raw Fresnel Phase difference (Section 4.3). After calibration, we will have the Fresnel Phase difference for every pair of subcarriers. Though the Fresnel Phase difference of any pair of subcarriers can be used to compute $\hat d$, more information often leads to higher accuracy and robustness against measurement and processing errors. Inspired by this, we propose a model fitting method to utilize the Fresnel Phase differences from all pairs of subcarriers. Model fitting would identify all variables in Equation \ref{e8} (Section 4.4) , including the object reflected path length, which determines the object's location in Fresnel zones.

In the next sections, we will present the three major components in details.
\subsection{Fresnel Phase Difference Calculation}
Given the amplitude time series $|H(\lambda, t)|$ for each subcarrier $\lambda$, the first step is to calculate the time shift $\Delta T_{ab}$ between each pair of subcarriers $(\lambda _a, \lambda _b)$. As there are up to $K$ subcarriers (K equals 30 in Intel 5300 NIC), there should be  $(_2^K) = (_2^{30}) = 435$ subcarrier pairs with their corresponding time shift $\Delta T_{ab}$, respectively. 

The time shift $\Delta T_{ab}$ of each subcarrier pairs is calculated by applying a similarity-based time shift estimation algorithm. The idea is to seek a parameter  $\Delta T$ such that the two amplitude time series $|H(\lambda_a, t)|$ and $|H(\lambda_b, t-\Delta T)|$ are alike with the highest correlation, i.e.,
\begin{equation}\label{e9}
\Delta T_{ab}=\mathop {\arg \max }\limits_{\Delta T \in w} [\rho (H({\lambda _a},t),H({\lambda _b},t - \Delta T))]
\end{equation}
where  $\rho (H({\lambda _a},t),H({\lambda _b},t-\Delta t))$ is the correlation in between and $w$ is the sliding window size.

The period of amplitude time series $H(\lambda, t)$ is computed by discrete Fourier transform (DFT) method. DFT decomposes the function into its constituent frequencies. The frequency with the highest amplitude has the strongest periodic strength, and the period $T$ is thus computed accordingly.


With the time shift $\Delta T$ and period $T$ obtained, the raw Fresnel Phase Difference is
\begin{equation}\label{e13}
\Delta \hat \varphi _{ab}=\Delta T_{ab}/T_{ab}
\end{equation}

It is worth noting that the computation of raw Fresnel Phase Difference $\Delta \hat \varphi _{ab}$ is also impacted by the sliding window size $w$. On the one hand, $\Delta T$ and  $\Delta \hat \varphi _{ab}$ obtained by Equation \ref{e9} and Equation \ref{e13} are the averaged values over the time window $w$. And thus a smaller $w$ is preferred to improve the computation accuracy. On the other hand, too small $w$ may have insufficient samples to compute the period $T$. In the extreme case when $w=1$, there is only one sample and the period can be any. We will investigate the impact of the parameter $w$ through empirical studies, and our experiment results show that $w=25$ samples per window achieves the highest localization accuracy.

\subsection{Phase Offset Calibration}
		In an indoor environment, NLoS paths will induce a fixed Fresnel Phase Offset $\varepsilon_{ab}$.  
		To compensate this offset, we have designed a phase offset calibration method. 
		Here, we first give the intuition of the calibration, then present the calibration process in  detail. We also compare the calibration process with fingerprint-based approaches to highlight the advantages of our approach.
		
		The basic idea of the phase offset calibration method is to measure the fixed offset in an offline manner and then calibrate the Fresnel phase difference by deducing the offset. Thus the critical point lies in measuring this offset. According to Equation \ref{e8}, the offset $\varepsilon_{ab}$ can be measured by Fresnel phase difference $\hat \Delta \varphi _{ab}$ minus $\Delta \varphi _{ab}$. $\hat \Delta \varphi _{ab}$ can be directly measured by the method proposed in the previous section. The question is how to get $\Delta \varphi _{ab}$.  Note that when an object moves along a predefined path, i.e., the corresponding $\hat d$ at these locations along the path are known in advance, then $\Delta \varphi _{ab}$ can be mathematically calculated using Equation \ref{e8}, since for a given LoS length $d_0$ and the two subcarriers with wavelength $\lambda_a$ and $\lambda_b$, $\Delta \varphi _{ab}$ is only determined by $\hat d$.
		
		In practice, to carry out the calibration process, we first setup the transceivers in the indoor environment, then we move a metal plate as a reflector along the perpendicular bisector of the two transceivers and collect the CSI magnitudes of all the subcarriers. We can then calculate the Fresnel phase offset by measuring the difference between the measured raw Fresnel phase difference $\Delta \hat \varphi _{ab}$ and the Fresnel phase difference $\Delta \varphi _{ab}$ derived in open space. As the locations and the corresponding $\hat d$ at these locations are known in advance, the Fresnel phase difference $\Delta \varphi _{ab}$ can be calculated in advance with Equation \ref{e6}. Therefore, the phase offset $ \varepsilon_{ab}$ can be obtained by Equation \ref{e8},\ref{e13} and \ref{e6}:
		\begin{equation}\label{e14}
		\begin{array}{lll}
		\varepsilon _{ab}&=&\Delta {T_{ab}}/{T_{ab}} - 2\pi ({\rm{\hat d -  }}{{\rm{d}}_0})(1/{\lambda _a} - 1/{\lambda _b})
		\end{array}
		\end{equation}
		Please note that the phase offset $\varepsilon_{ab}$ is associated with each pair of subcarriers. For each measurement, we actually obtain a phase offset matrix  $({\varepsilon _{ab}}) \in \Re {^{K \times K}}$, where $K$ is the number of subcarriers.
		
		Compared with fingerprint-based approaches, which require the collection of CSI signals at each location in the entire sensing area to create the fingerprint map to infer the subject's location, our calibration process requires very little sampling along one path with several locations. As such, MFDL has much lower overhead than those methods. 

\subsection{Model Fitting for FPM}

\begin{figure}[!t]
	\centering
	\includegraphics[width=3.5in]{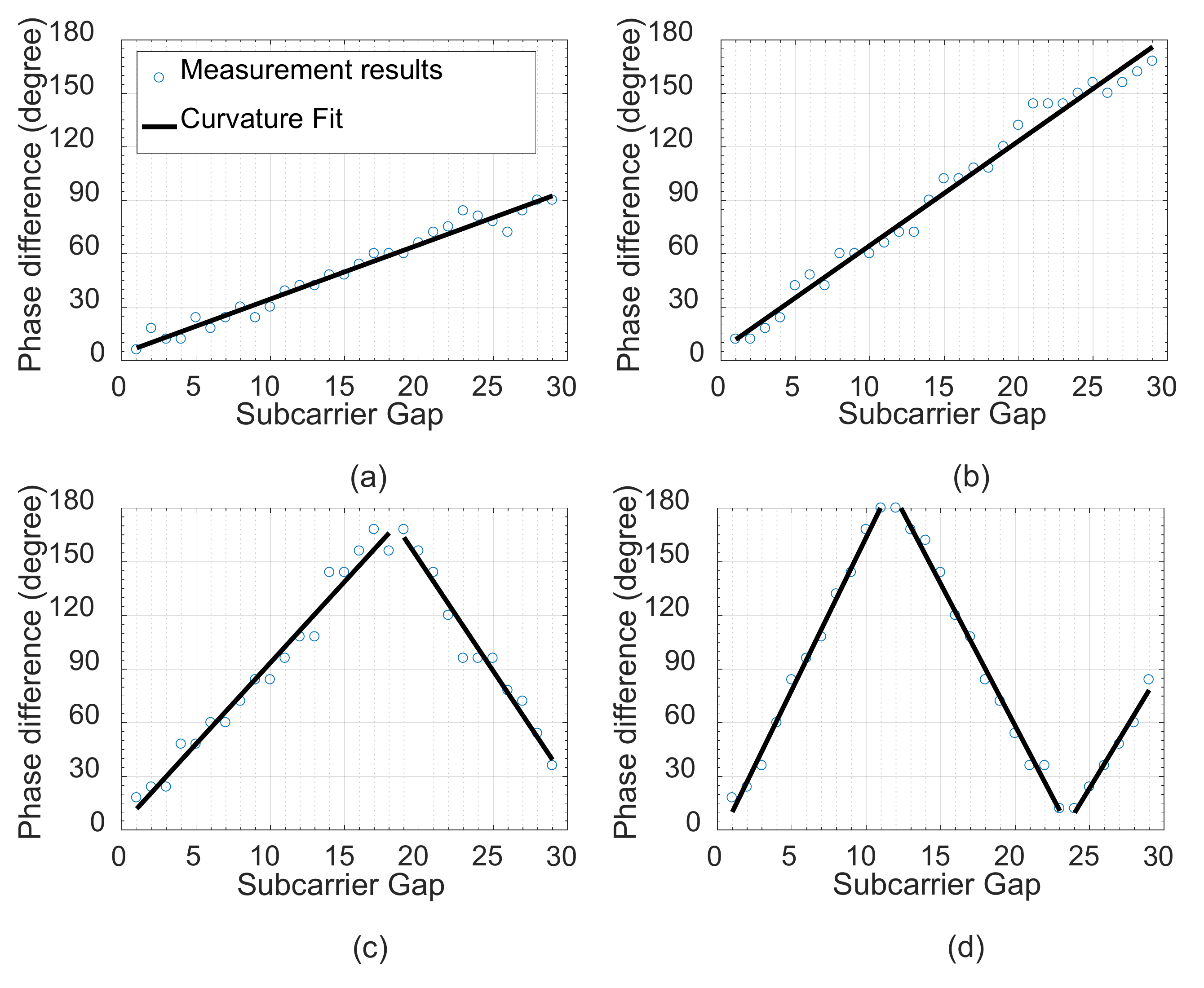}
	\caption{Illustrative examples of Modeling Fitting for FPM.}
	\label{Fig551}
\end{figure}
\begin{figure*}[!t]
	\centering
	\includegraphics[width=6.6in]{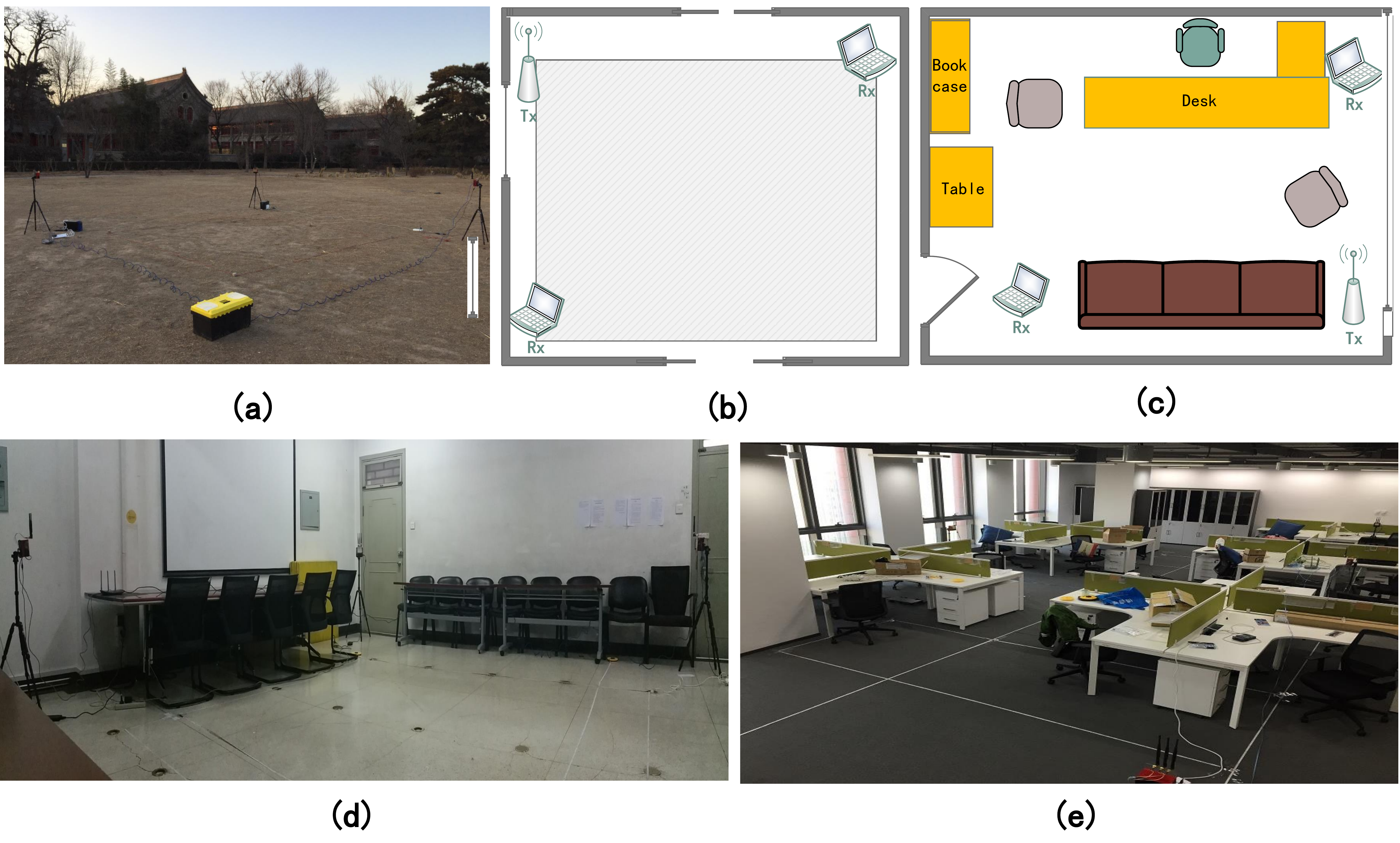}
	\caption{Five different environments used for our experiments: (a) outdoor, (b) hall, (c) office room, (d) meeting room, (e) student room.}
	\label{Fig621}
\end{figure*}
Ideally, the reflected path length $\hat d$ can be computed by Equation \ref{e6} with any pair of subcarriers. In practice, there are always errors introduced from the CSI measurements $|H(\lambda,t)|$, the process computing $\Delta T$ and $T$, and calibration error to estimate $\varepsilon_{ab}$. To minimize the error, we apply all subcarrier pairs and adopt a least square fitting algorithm to identify $\hat d$.
The rationale behind is that according to Equation \ref{e6}, the Fresnel phase difference $\Delta \varphi_{ab}(\hat d)$ not only has a linear relationship with the resided Fresnel zones if given a pair of subcarriers, but also has a linear relationship with the frequency gap of the subcarrier pairs if given $\hat d$. Therefore, it is safe to assume the reflected path length $\hat d$ remains the same during a very short sliding window (0.05s in our system). Thus based on this observation, we could apply the standard linear least square method for the model fitting.


Fig. \ref{Fig551} shows some model fitting examples. While the x-axis is the frequency gap $\Delta f$ in terms of subcarrier index, the y-axis is the Fresnel phase difference. For example, $x=5$ means the gap between two subcarriers are 5, e.g. with subcarrier 1 and 6, subcarrier 2 and 7, and so on. 

Notice that when the object is sufficiently far away from the transceivers and the frequency gap is large enough, e.g., Fig. \ref{Fig551} (c) $gap>20$ , the Fresnel Phase difference can be over $\pi$. And for Fig. \ref{Fig551} (d) $gap>25$, the Fresnel Phase Difference could be more than $2\pi$. In such cases, the fitting curves are zig-zag fold lines. As we have no idea in advance about the number of folds, what we can do is to try all the options, e.g., a line with no fold, single fold and two folds, and select the reflected path length $\hat d$ with the minimal fitting error. After the reflected path length $\hat d$ is obtained, the location in the corresponding Fresnel zones is known, so is the intersection area in the 2-dimensional plane.

\section{MFDL Performance Evaluation}
\label{sec:eval} 

In this section, we evaluate the performance of our MFDL system for device-free localization. We first describe the system implementation and experimental setup. We then present detailed experimental results covering the overall localization performance, system robustness under diverse scenarios, as well as the impact of individual system design components. 

\subsection{System Implementation}

We have implemented MFDL using an 802.11n Wi-Fi network consisting of one Wi-Fi transmitter and three Wi-Fi receivers. We use four Gigabyte BXi3H-5010 Brix mini-PCs with Intel 5300 wireless NIC. Each NIC is equipped with external omni-directional antennas and we only choose one antenna to receive or transmit packets. Each mini-PC has 2G memory and runs Ubuntu 14.04 LTS. The three receivers collect CSI data using tools developed by Halperin et al \cite{halperin2011tool}, and pass the data to a backend server for processing. The transmitter is configured to send packets in injection mode using the default transmission power setting. Since only one transmitter is needed in our experiments, there is no need to use multiple transmission bands in order to avoid interference among multiple transmitters. The transmitter drops some packets every 10 seconds in a predefined pattern to act as a sync signal. The three receivers can therefore align with each other according to this signal. All four devices are mounted on tripods and the antennas are positioned at 1.5m above ground, similar to the setup of prior works~\cite{wang2016lifs, li2016dynamic, wu2016widir}  for proper detection of human body.

The system is configured to run in the 5GHz frequency band with 40MHz bandwidth. By using the higher frequency band (instead of 2.4GHz) and wider bandwidth (instead of 20MHz), finer-grained Fresnel phase difference between subcarriers can be captured, thus allowing for more precise Fresnel zone separation and target localization. 
Typical indoor walking speed is between 0.5m and 1.5m per second, which corresponds to  about 10Hz to 70Hz fluctuation in the 5GHz frequency band based on the FPM model. Therefore, we set the sampling rate to 500 packets per second, which is sufficient to capture such fluctuations.
\begin{figure}[!t]
	\centering
	\includegraphics[width=3.5in]{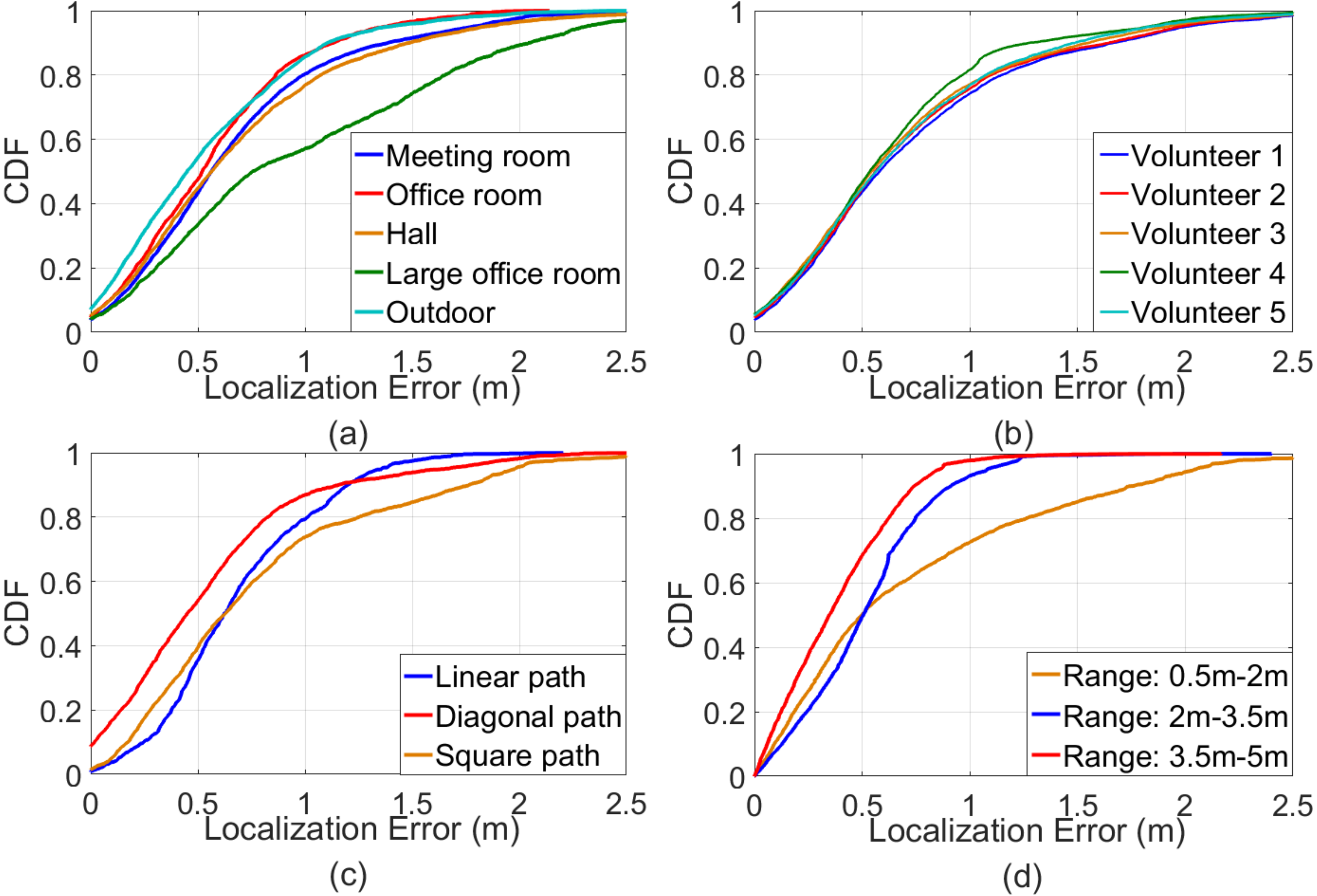}
	\caption{MFDL performance for different environments (a), different human subjects (b), different types of paths (c), and different sensing ranges (d).}
	\label{51new}
\end{figure}
\subsection{Experimental Setup}

As shown in Fig.~\ref{Fig621}, we have conducted our experiments in five different environments: 
\begin{itemize} 
	\item (a) Outdoor: An outdoor open space with a $6m \times 6m$ sensing area
	\item (b) Hall: An empty hall of size $7m\times7m$  and $5m \times 5m$ sensing area 
	\item (c) Office room: An office room of size $3m \times 4m$ with one sofa, two tables, and one bookcase
	\item (d) Meeting room: A meeting room of size $6m \times 6m$ and $5m \times 5m$ sensing area with a long meeting table and dozens of chairs
	\item (e) Student room: A student room of size $13m \times 7m$ and $12m \times 6m$ sensing area with many chairs, tables, and other cluttered objects
\end{itemize} 

For the first four environments, three transceivers were positioned at three of the corners (Figure~\ref{Fig621}(b)). For the student room, four transceivers were used, 
		with one transmitter positioned at 6m along the long edge, two receivers positioned at 0m  and 12m along the same long edge, and one receiver positioned at 6m on the opposite long edge (i.e., each receiver is 6m away from the transmitter). Please note that due to structural constraints,  the sensing areas for the hall, meeting room, and student room are smaller than the actual room size. As shown in the figure, most of the rooms have furniture such as tables, chairs, bookcase, and other cluttered objects, which are common in indoor environments. We chose test locations that were spaced at 0.5m in each sensing area. The true location coordinates were measured using a Bosch GLM-80 laser range-finder and marked in the sensing area beforehand. We recruited 5 volunteers (four male and one female; age: 21--32 years; height: 1.6--1.83m)  to perform the experiments. Each volunteer was instructed to walk along a predefined path\footnote{Our empirical observations using arbitrary paths showed similar performance. We only report results of predefined paths here, since it is difficult to obtain accurate ground truth for arbitrary paths and our system by design depends only on the penetration of Fresnel zones and not the actual path.} We experimented with different types of paths in each environment, including straight lines that are horizontal, vertical, or diagonal to the transceivers, as well as rectangle, square, and diamond paths. Each path was repeated at least 5 times under each environment. In total, we experimented with 782 paths for the five environments, providing a comprehensive and roughly even coverage of all the test locations in the sensing areas. 

{\bf Performance metric.} We use localization error to measure the performance of our system. Since typical human body has a width of around 40cm, we treat a human target as a cylinder object instead of a point. Therefore, a location estimation is considered to be correct (i.e., zero error) if it falls within 20cm of the true location. Otherwise, the localization error is calculated as the minimum distance between the estimated location and the edge of the cylinder. Please note that this is the same performance metric used by LiFS \cite{wang2016lifs} and DynamicMUSIC \cite{li2016dynamic}. We use the cumulative distribution function (CDF) and median error to measure the aggregated localization error for each experimental setting.

\subsection{Overall Localization Performance} 

Fig.~\ref{51new}(a) shows the CDF of localization error under the five different environments. Overall, MFDL performs well in all five environments. Using only three transceivers, MFDL achieves 45cm median error for the outdoor open space, and 50--60cm median error in the first three indoor environments (i.e., hall, office room, and meeting room). When increasing the number of transceivers to four, MFDL achieves 75cm median error for the large student room. In comparison, the state-of-the-art LiFS achieved 1.1m median error in a 70$m^2$ area using 11 transceivers.

As can be seen in Fig.~\ref{51new}(a), MFDL also performs consistently well for three of the indoor environments, even though they have different sizes and very different multipath propagation characteristics. To further analyze the robustness of MFDL under different scenarios, we also plot in Fig.~\ref{51new} the performance of MFDL for different human subjects, different types of paths, and different sensing ranges. We can see in Fig.~\ref{51new}(b) that MFDL achieves similar performance for all five volunteers. For the different types of paths, MFDL performs better for the diagonal paths (45cm median error). This is because, when walking along the diagonal paths, the human target penetrates Fresnel zones more effectively in both dimensions, making it easier to locate the human target. Still, MFDL performs reasonably well for linear and square paths with 63cm median error as shown in Fig.~\ref{51new}(c). Finally, as shown in Fig.~\ref{51new}(d), as the sensing range (i.e., sensing distance to the human target) increases, MFDL's performance does not degrade.

{\bf System robustness:} Based on the results in Fig.~\ref{51new}, we can see that MFDL performs consistently well for different environments, different human subjects, different types of paths, and different sensing ranges. Such robustness is important for real-world applications and demonstrates the high potential of MFDL for practical deployment. This robustness is the benefit of the underlying FPM model that MFDL is based on, which is a generic model capable of capturing the precise relationship between target location and CSI under diverse scenarios. The individual design components we propose in the MFDL system also contribute to the overall system performance, which we evaluate next. 

\subsection{Individual System Design Performance}  

Here, we consider individual system design features and evaluate their contributions to the overall system performance of MFDL. Specifically, we evaluate the effectiveness of curvature fitting and phase offset calibration, as well as the impact of sliding window size. 

\begin{figure}[!t]
	\centering
	\includegraphics[width=3.5in]{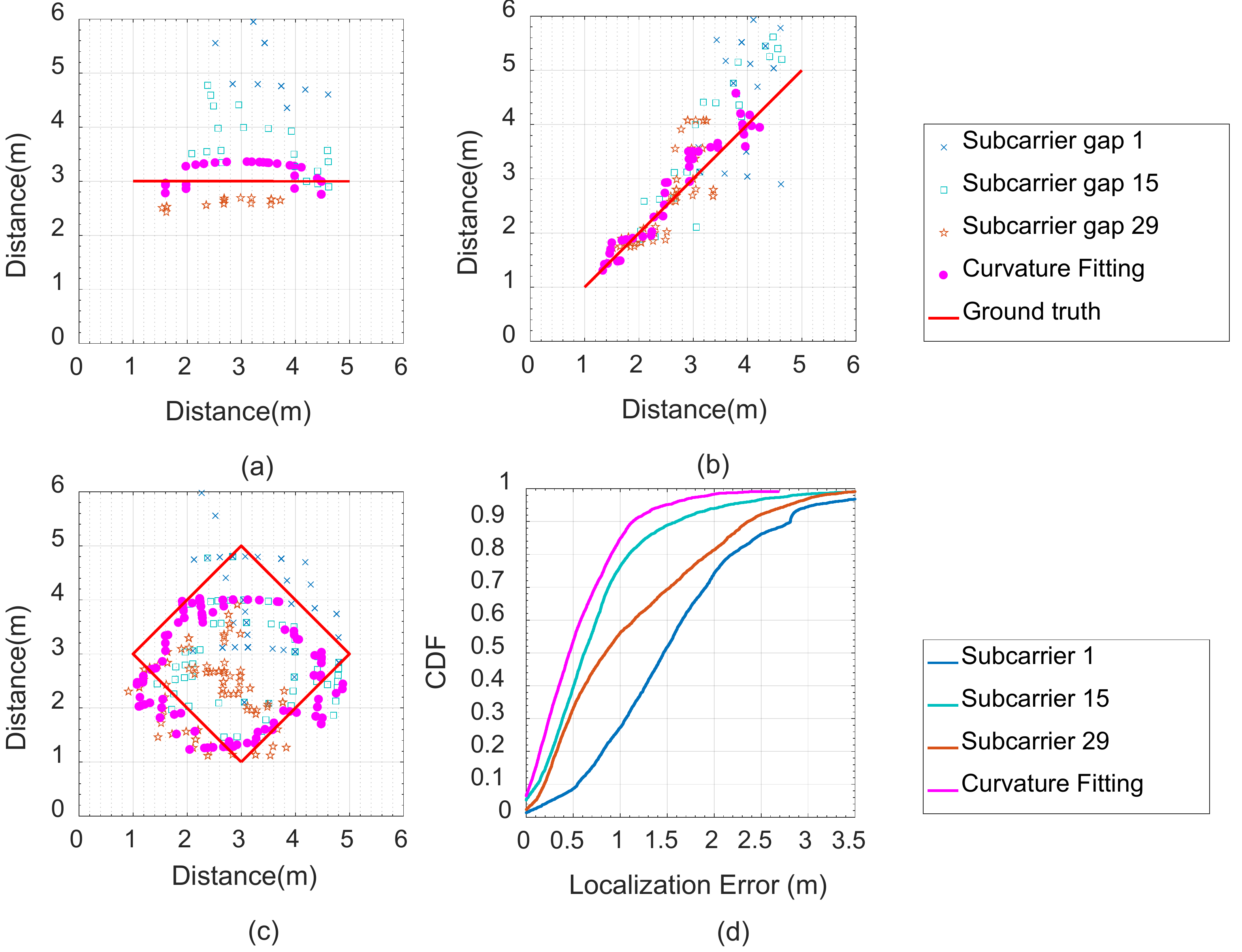}
	\caption{MFDL performance using different groups of subcarriers with fixed gaps, and the use of curvature fitting.}
	\label{52new}
\end{figure}

\subsubsection{Effectiveness of Model Fitting (Curvature Fitting)}
The Intel 5300 wireless NIC reports 30 CSI values for the 40MHz bandwidth, which means that each CSI subcarrier is spaced in 4 $\times$ 0.3125MHz. According to our FPM model, given a pair of subcarriers with certain frequency gap , we can estimate the location based their Fresnel zone phase difference. Instead of choosing subcarriers with a fixed frequency gap, we propose a curvature fitting process in MFDL to determine optimized combinations of subcarriers. Here, we evaluate the impact of different frequency gaps as well as the effectiveness of our proposed curvature fitting approach. We conducted  experiments in the outdoor open space. We  roughly divided all possible subcarrier pairs into three frequency gap groups: gap 1, gap 15, and gap 29. For each group, we calculate the median value of the localization errors of all the subcarrier pairs in the group. We then compare the results of different frequency gap groups with our proposed curvature fitting approach. 
Fig.~\ref{52new} shows the comparison results. We can see that our proposed curvature fitting approach consistently outperforms the other approaches and achieves a median error of around 45cm in the outdoor environment.

\begin{figure}[!t]
	\centering
	\includegraphics[width=3.5in]{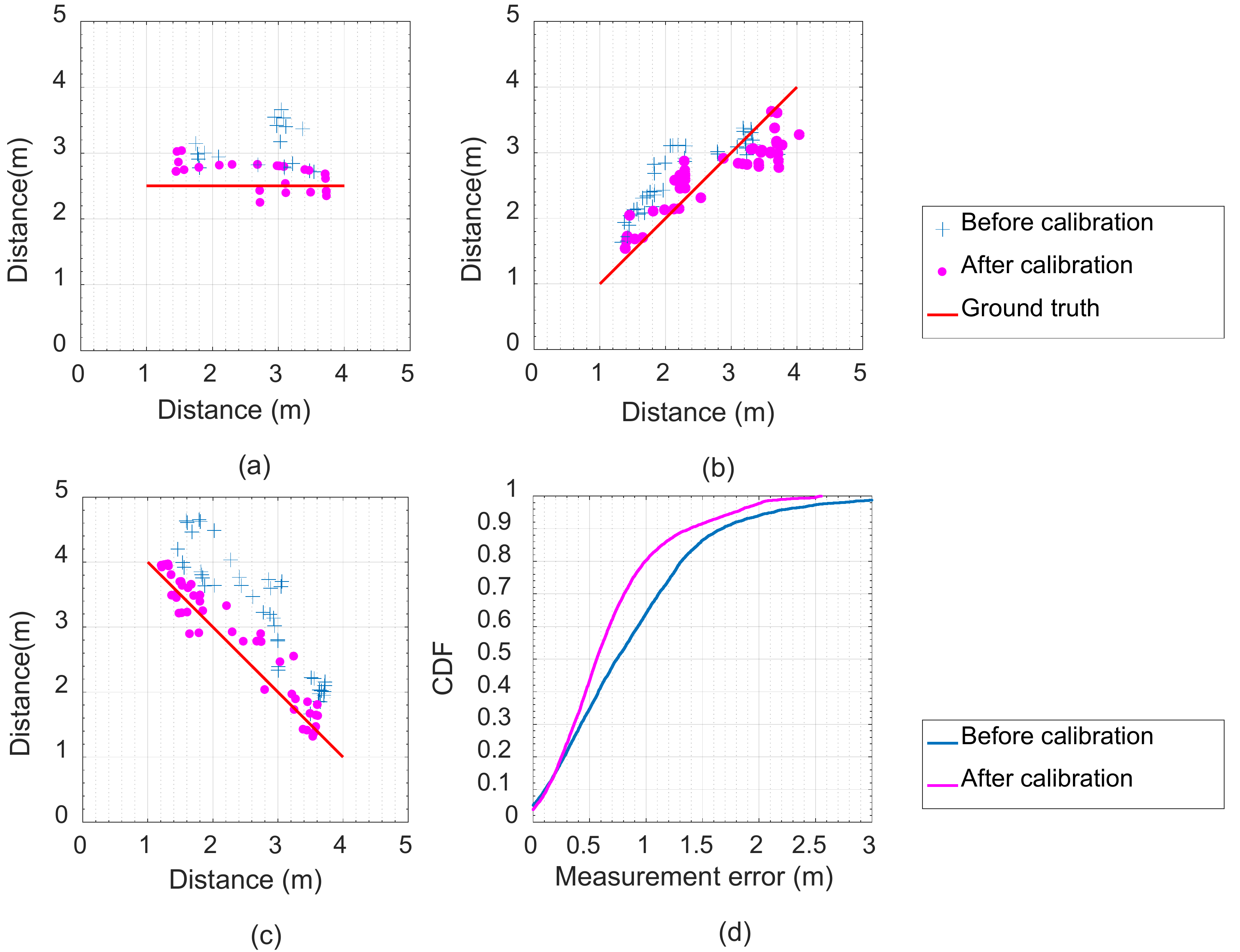}
	\caption{MFDL performance before or after phase offset calibration.}
	\label{53new}
\end{figure}
\begin{figure}[!t]
	\centering
	\includegraphics[width=3.5in]{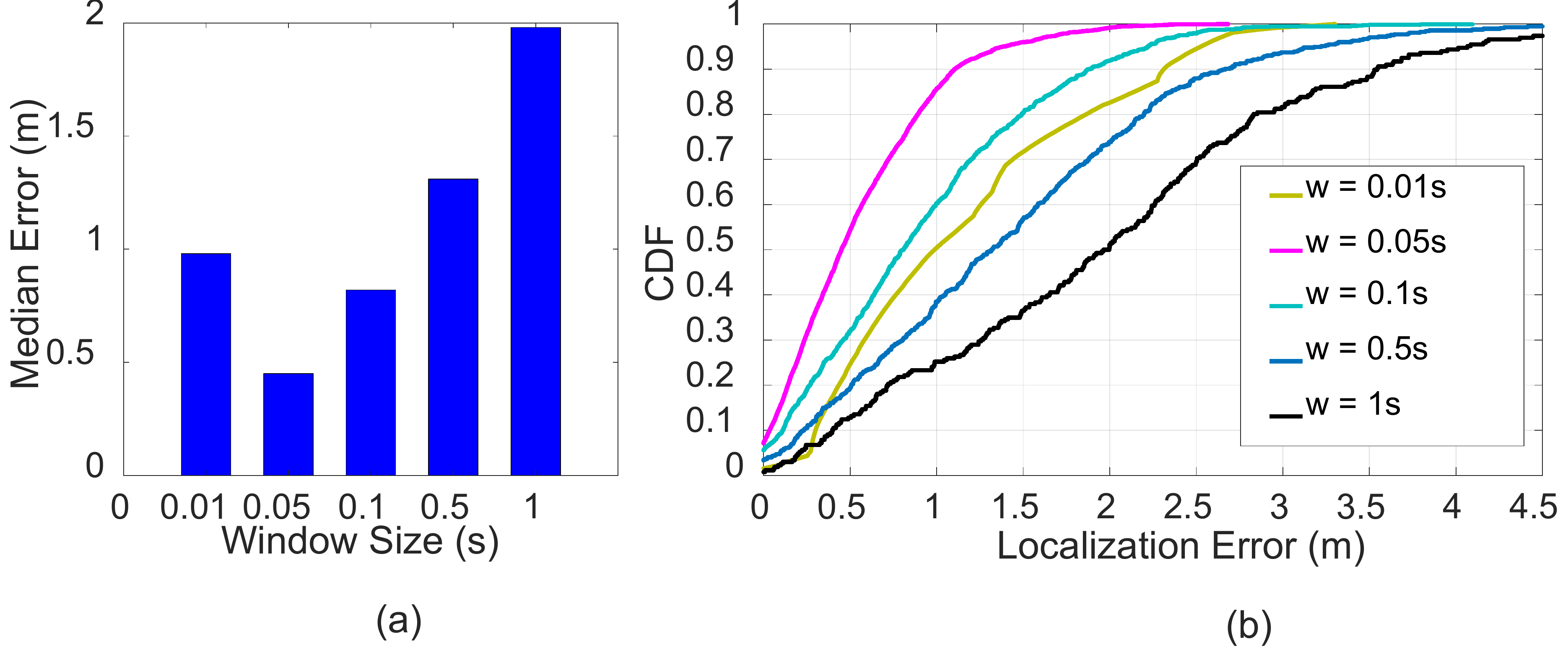}
	\caption{MFDL performance using different sliding window sizes.}
	\label{54new}
\end{figure}
\subsubsection{Effectiveness of Phase Offset Calibration}
In Section 3.4.2, we have verified our FPM model Equation \ref{e8} with perfect reflector in the indoor space. For real-world use of the MFDL system, we propose a calibration process to correct phase offset. To demonstrate the effectiveness of our phase offset calibration method, we conducted experiments with real human subjects in the indoor environments. Fig.~\ref{53new} shows the localization results before and after calibration and we can see that for different test locations and different paths, the localization performance improves significantly with the use of phase offset calibration, reducing the median error from about 73cm to only 55cm.

\subsubsection{Impact of Sliding Window Size}
For the initial measurement of Fresnel phase difference, we use the CSI signals in a specific sliding window for the time shift calculation and cycle estimation. This sliding window size can impact the overall localization performance. Intuitively, a smaller window size makes our system more sensitive to location change in time series. However, too small a window size also makes it difficult to calculate time shift and estimate cycle. To evaluate how this sliding window size impacts our localization performance, we conducted experiments in the outdoor open space with window sizes ranging from 0.01s to 1s. The results shown in Fig.~\ref{54new} match our expectations that a proper window size should not be too narrow or too wide. Empirically, the 0.05s sliding window size is a good choice. Since the typical walking speed is about 0.5-1.5m/s, a 0.05s window size means that we can expect 1$\sim$3 CSI cycles in the sliding window, during which period the human target  would have moved only a small distance of 2.6$\sim$7.8cm.

\section{Conclusion and Future Work}
\label{sec:conclusion} 

In this work, we study device-free localization using COTS Wi-Fi technology. We explore the Fresnel Zone Theory and propose the Fresnel Penetration Model. FPM is a generic model that captures the linear relationship between specific Fresnel zones and multicarrier Fresnel phase differences, as well as the Fresnel phase offset caused by multipath environment. We validate the goodness of fit of FPM in both open outdoor and complex indoor environments. Guided by FPM, we propose MFDL, a multicarrier FPM based localization system, which consists of a number of signal processing methods to address some practical challenges with CSI measurements, particularly the Fresnel phase difference estimation and phase offset calibration in multipath-rich indoor environments. Experimental results  show that using only three transceivers, the median error is only 45cm for a 36$m^{2}$ outdoor environment, and the median errors are 0.55m, 0.52m, and 0.55m, for three typical indoor environments, respectively. Increasing the number of transceivers to 4 enables us to achieve 75cm median localization error for a 72$m^2$ area, compared with the 1.1m median error achieved by the state-of-the-art LiFS with 11 transceivers (4 APs and 7 clients) in a 70$m^2$ area. 

FPM is a generic model for device-free localization. In this paper, we show how to apply it for single object localization for both indoor and outdoor environments. As our future work, there are several issues to investigate. First, although many real-world applications exist for single-user localization, such as assistive living for aged or disabled people who (temporarily) live alone, multi-object localization is still a high-priority problem that we would like to investigate in the future. Second, in our current MFDL system design, we apply an offline method to calibrate CSI phase offset. This calibration overhead can be large for highly dynamic environments. Third, our current evaluation focuses on localization in a single room, which we plan to expand to multiple rooms such as  apartments in the future.  And last but not least, given the noisy distortion of CSI measurements in real-world environments, more intelligent signal processing methods that build on top of our FPM model and MFDL system would be desirable. 

\emph{The demo video for our device free localization system MFDL can be found via the following permalink:} \url{http://wanghao13.top/wordpress/index.php/2017/07/21/mfdl/}.

\ifCLASSOPTIONcompsoc
  \section*{Acknowledgments}

Daqing Zhang is the corresponding author. This research is supported in part by the National Key Research and Development Plan Grant No. 2016YFB1001200, and the U.S. National Science Foundation under grant No. 1442971.

\else
  \section*{Acknowledgment}
\fi
\ifCLASSOPTIONcaptionsoff
  \newpage
\fi



%
\bibliographystyle{IEEEtran}
\bibliography{bib2}
%

\begin{IEEEbiography}[{\includegraphics[width=1in,height=1.25in,clip,keepaspectratio]{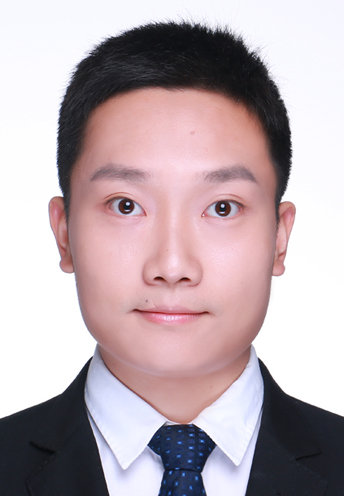}}]{Hao Wang}
received his Ph.D. degree in computer science from Peking University in 2017. He is going to be a research engineer for the Huawei Technologies Co. Ltd. His research interests include mobile crowdsensing and ubiquitous computing.
\end{IEEEbiography}

\begin{IEEEbiography}[{\includegraphics[width=1in,height=1.25in,clip,keepaspectratio]{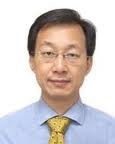}}]{Daqing Zhang}
received his Ph.D. from University of Rome “La Sapienza” in 1996. He is a Chair Professor at School of EECS, Peking University, China. His research interests include context-aware computing, urban computing, mobile computing, big data analytics, pervasive elderly care, etc..  He has published more than 200 technical papers in leading conferences and journals. He served as the general or program chair for more than 10 international conferences, giving keynote talks at more than 16 international conferences. He is the associate editor for ACM Transactions on Intelligent Systems and Technology, IEEE Transactions on Big Data, etc.. He is the winner of the Ten-years CoMoRea impact paper award at IEEE PerCom 2013, the Honorable Mention Award at ACM UbiComp 2015 and 2016, the Best Paper award at IEEE UIC 2015 and 2012.
\end{IEEEbiography}
\vfill
\begin{IEEEbiography}[{\includegraphics[width=1in,height=1.25in,clip,keepaspectratio]{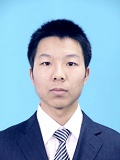}}]{Kai Niu}
 received the MS degree in Computer Technology from School of Electronic and Information Engineering, Xi’an Jiaotong University, in 2016. He is currently working toward the PhD degree in computer science with School of Electronics Engineering and Computer Science, Peking University. His research interests include ubiquitous computing and mobile computing. 
\end{IEEEbiography}

\begin{IEEEbiography}[{\includegraphics[width=1in,height=1.25in,clip,keepaspectratio]{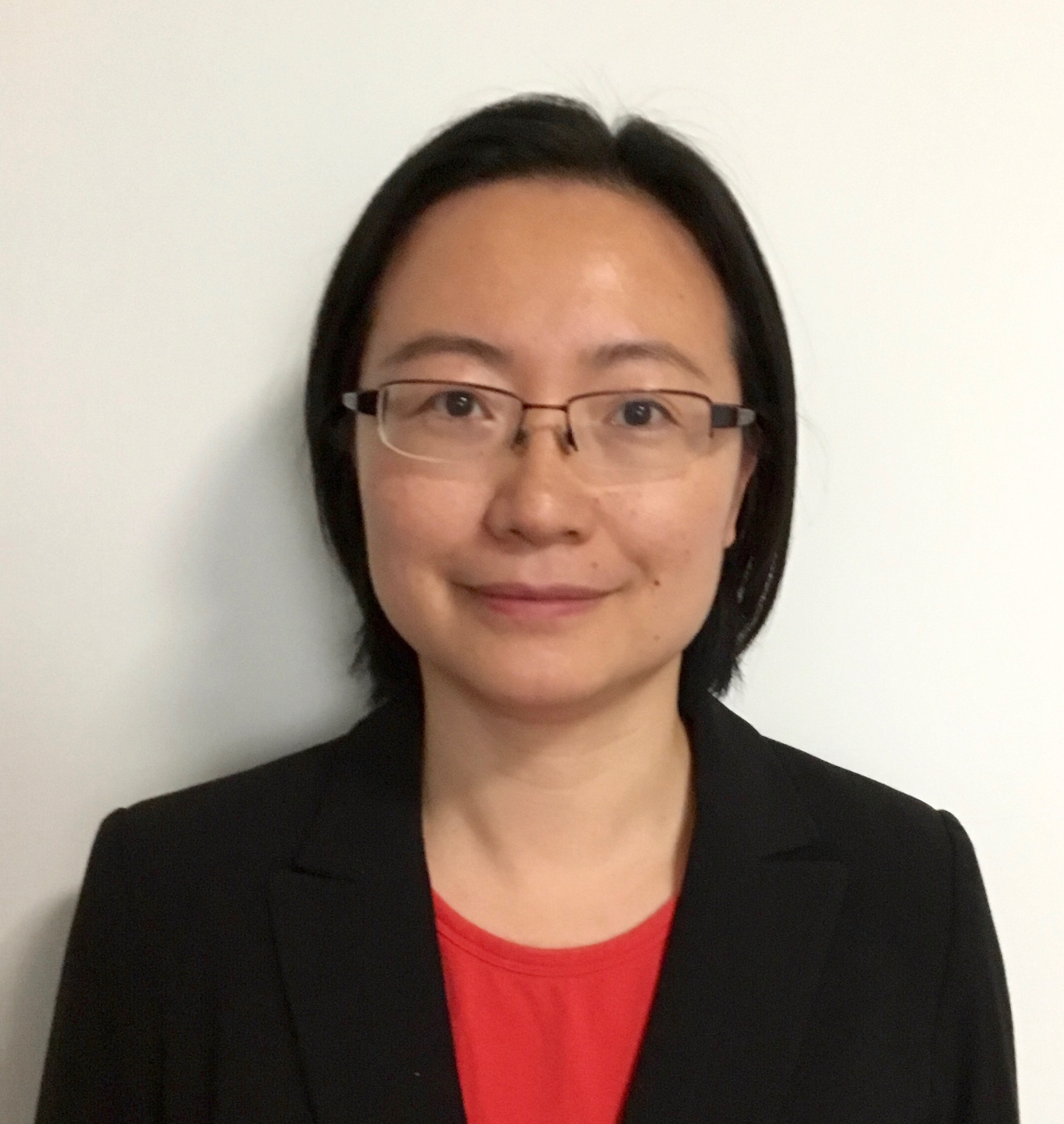}}]{Qin Lv}
	Qin Lv received her PhD degree in computer science from Princeton University in 2006. She is an associate professor in the Department of Computer Science, University of Colorado Boulder. Her main research interests are data-driven scientific discovery and ubiquitous computing. Her research spans the areas of multi-modal data fusion, spatial-temporal data analysis, anomaly detection, mobile computing, social networks, recommender systems, and data management. Her research is interdisciplinary in nature and interacts closely with a variety of application domains including environmental science, Earth sciences, renewable and sustainable energy, materials science, as well as the information needs in people`s daily lives. Lv is an associate editor of PACM IMWUT, and has served on the technical program committee and organizing committee of many conferences. Her work has more than 4,000 citations.
\end{IEEEbiography}
\vfill
\vfill

\begin{IEEEbiography}[{\includegraphics[width=1in,height=1.25in,clip,keepaspectratio]{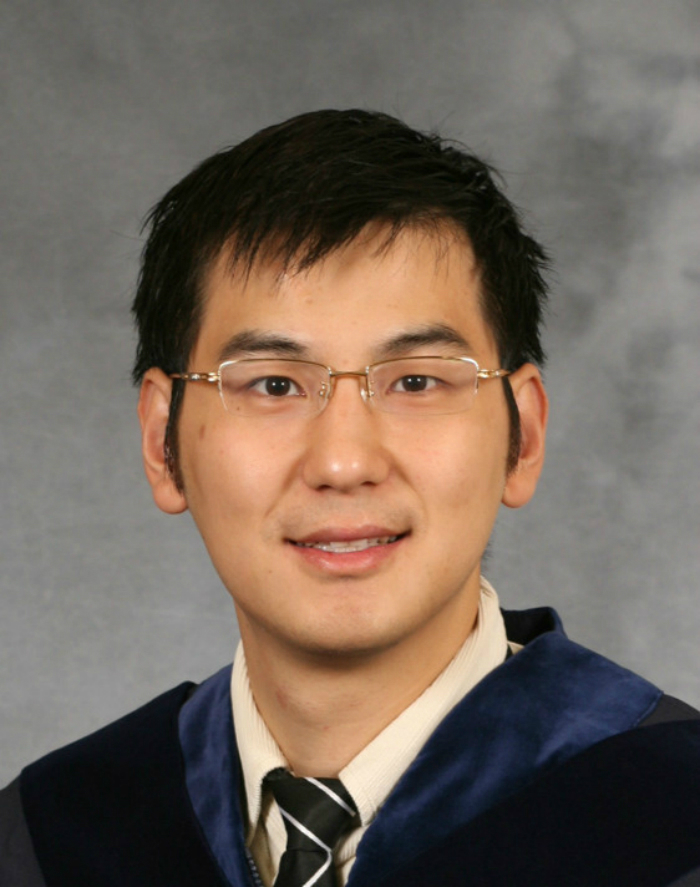}}]{Yunhuai Liu}
	is now an associate professor with Peking University, P.R. China. He received his PhD degree in Computer Science and Engineering from Hong Kong University of Science and Technology (HKUST) in 2008.  From 2008 to 2010 he worked in HKUST as a research assistant professor. In the year 2010, he joined Shenzhen Institutes of Advanced Technology, Chinese Academy of Sciences as an associate professor. From 2011 to 2016, he was with the Third Research Institute of Ministry of Public Security, China. His research interests include wireless sensor networks, cognitive radio networks, and pervasive computing.
	
\end{IEEEbiography}

\begin{IEEEbiography}[{\includegraphics[width=1in,height=1.25in,clip,keepaspectratio]{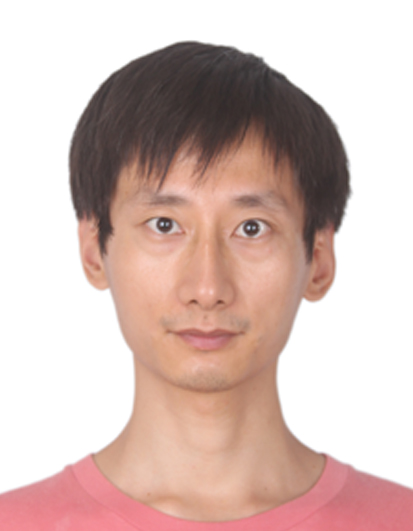}}]{Dan Wu}
	is a PhD student in computer science in the School of Electronics Engineering and Computer Science at Peking University. His research interests include software modeling and ubiquitous computing. Wu received a BS in computer science from the University of Science and Technology of Beijing.
\end{IEEEbiography}

\begin{IEEEbiography}[{\includegraphics[width=1in,height=1.25in,clip,keepaspectratio]{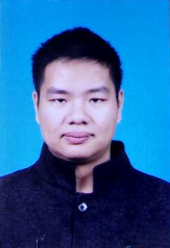}}]{Ruiyang Gao}
	received the BE degree in Computer Science and  Technology from	Shandong University,Jinan,China in 2016. He is currently	working toward the PhD degree in computer science	in the School of Electronics Engineering	and Computer Science, Peking University. His research interests include ubiquitous computing and mobile computing.
\end{IEEEbiography}

\begin{IEEEbiography}[{\includegraphics[width=1in,height=1.25in,clip,keepaspectratio]{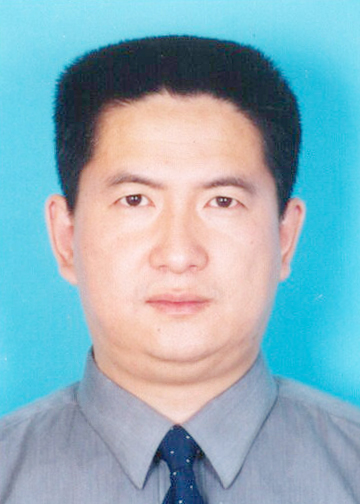}}]{Bin Xie}
received the Ph.D degree in Computer Science from National University of Defense Technology of China in 1998. He is a professor in the school of EECS, Peking University since 2007. He has published more than 80 papers in leading conferences and journals, such as FSE, POPL, UbiComp, etc. He was awarded the Distinguished Young Scholars of NSFC as his contribution in the domain of software reuse. His research interests include software engineering, formal methods and distributed and reactive system.
\end{IEEEbiography}

\vfill


\end{document}